\documentclass[twocolumn]{aastex631}
\definecolor{phoebe}{rgb}{0.7, 0.0, 0.5}

\usepackage{amsmath, mathtools}
\usepackage[T1]{fontenc}
\usepackage{physics}
\usepackage{import}

\newcommand{\revise}[1]{#1}
\newcommand{\reviseTwo}[1]{#1}
\newcommand{\reviseThree}[1]{#1}

\newcommand{\Nemb}{\ensuremath{N_{\mathrm{emb}}}} 
\newcommand{\Nsp}{\ensuremath{N_{\mathrm{small}}}} 
\newcommand{\Nsg}{\ensuremath{N_{\mathrm{sur,giant}}}}

\graphicspath{{./}{figures/}}

\begin{document}

\title{Effects of Outer Giant Planets on In Situ Formation of Inner Super-Earths}

\correspondingauthor{Phoebe Sandhaus}
\email{pjs5535@psu.edu}

\author[0000-0002-6019-4818]{Phoebe Sandhaus}
\affiliation{Department of Astronomy \& Astrophysics, 525 Davey Laboratory, The Pennsylvania State University, University Park, PA 16802, USA}
\affiliation{Center for Exoplanets and Habitable Worlds, Pennsylvania State University, University Park, PA, 16802, USA}

\author[0000-0001-9677-1296]{Rebekah I. Dawson}
\affiliation{Department of Astronomy \& Astrophysics, 525 Davey Laboratory, The Pennsylvania State University, University Park, PA 16802, USA}
\affiliation{Center for Exoplanets and Habitable Worlds, Pennsylvania State University, University Park, PA, 16802, USA}

\author[0000-0003-2372-1364]{Mariah G. MacDonald}
\affiliation{Department of Physics, The College of New Jersey, 2000 Pennington Road, Ewing, NJ 08628, USA}

\author[0000-0003-4709-2689]{Cody J. Shakespeare}
\affiliation{Department of Physics and Astronomy, University of Nevada, Las Vegas, 4505 S. Maryland Pkwy, Las Vegas NV 89154, USA}

\author[0000-0002-2432-833X]{Sarah Morrison}
\affiliation{Department of Physics, Astronomy, \& Materials Science, Missouri State University, Springfield, MO 65897, USA}

\shorttitle{DRAFT}
\shortauthors{Sandhaus}

\begin{abstract}
Recent studies have found an observational correlation between the presence of outer giant planets and inner super-Earths\revise{, which implies} that outer giants do not suppress the formation of super-Earths.  We simulate late-stage in situ planet formation in the presence of outer giant planets \revise{using $N$-body simulations}.  We investigate the effects of two sets of outer giants: the four Solar System giant planets and three dynamically active giant planets.  Compared to systems without outer \revise{giants}, we find that systems with the Solar System giants tend to form inner super-Earths that are more compact, coplanar, and circular, while the systems with the dynamically active giants form inner super-Earths that are more eccentric, inclined, and widely spaced, with lower intrinsic multiplicity. Including a contribution from systems that form with dynamically active giant planets allows us to match observable quantities of super-Earths, including their two component eccentricity distribution. However, matching the observed population requires different formation conditions prior to the giant impact stage for systems with vs.\ without giant planets. In our model, observed super-Earths that form in the presence of dynamically active outer giants emerge from disks with lower solid surface densities and without a depleted gas stage, suggesting that the giant planets may have reduced---but not prevented---delivery and/or accretion of solids in the inner disk.  With a large enough sample of inner and outer systems, we could break down occurrence rates of inner super-Earths based on the properties of outer giants, and vice versa, and then compare these conditional probabilities with simulations.  

\end{abstract}


\section{Introduction} \label{sec:intro}

In the Solar System, many aspects of how the outer giant planets affect the inner terrestrial planets, especially Earth, have been extensively studied. Initially, Jupiter was thought to act as a gravitational shield against cometary impacts \citep{Wetherill1994_Ap&SS}, but more recent studies have found that Jupiter's presence could enhance the impact flux on the terrestrial planets \citep{HornerJones2009_IJAsB, Grazier2016_AsBio} and deliver volatiles to early Earth \citep{Carter-BondO'Brien2014_IAUS, O'BrienWalsh2014_Icar, Grazier2016_AsBio}. It is long-established that the long-term climate variability of Earth is driven by orbital variations, also called Milankovitch cycles \citep{Kerr1978_Sci}, which are dominated by Jupiter's gravitational influence \citep{HornerWaltham2014_arXiv}.  Dynamical simulations that implement small changes to Jupiter's orbit find that long-term climatic variations of Earth are significantly altered \citep{HornerGilmore2015_arXiv}. \revise{Other simulations have sought to reproduce the planetary properties of the terrestrial planets in the Solar System by varying the orbital configuration of Jupiter and Saturn.  When Jupiter and Saturn were put on circular orbits, the resulting Mars-analog was too massive; whereas, when the gas giants were given slightly higher eccentricities, the simulations produced an accurately-sized Mars, but lacked the ability to deliver water to early Earth \citep{RaymondO'Brien2009_Icar}.} These results demonstrate that even slight changes in the architecture of outer giant planets can have considerable effects on the \revise{orbital and physical properties, and thus the habitability, of our Solar System's terrestrial planets}. 

\revise{Beyond our Solar System,} we have observed many instances of outer giant planets in the same system as close-in super-Earths \citep[e.g.,][]{FischerMarcy2008_ApJ, WrightUpadhyay2009_ApJ, SantosSanterne2016_A&A, DedrickFulton2021_AJ, AzevedoSilvaDemangeon2022_A&A, Lillo-BoxGandolfi2023_A&A}.
Occurrence rates using both transit and radial velocity data indicate the presence of an outer giant planet does not decrease -- and may even increase -- the likelihood of finding an inner super-Earth (\citealt{ZhuWu2018_AJ, BryanKnutson2019_AJ, FultonRosenthal2021_ApJS, RosenthalKnutson2022_ApJS}\revise{; \citealt{BonomoDumusque2023_A&A}}). Recently, the California Legacy Survey \citep{RosenthalFulton2021_ApJS_CLS}, a blind radial velocity survey of 719 stars lasting over three decades, was used to explore the likelihood of inner super-Earths accompanying outer giant planets.  \citet{RosenthalKnutson2022_ApJS} find that the probability of a system hosting an inner super-Earth \revise{(0.023--1 AU, 2--30 $M_\oplus$)} given an observed Jupiter analog (3--7 \revise{AU}, 0.3--13 \revise{$M_\text{Jup}$}) is $0.32^{+0.24}_{-0.16}$, compared to the overall probability of $0.276^{+0.058}_{-0.048}$ for inner super-Earths. Conversely, the probability of a system hosting a Jupiter analog given an observed inner super-Earth is $0.133^{+0.097}_{-0.063}$, compared to the overall probability of $0.072^{+0.014}_{-0.013}$ for Jupiter analogs.  \reviseThree{Another study using the California Legacy Survey found that the super-Earth occurrence rates are only enhanced for Saturn-size ($M< 1.5$ $M_\text{Jup}$) outer giant planets and not for super-Jupiters \citep[$M \geq 1.5$ $M_\text{Jup}$,][]{lefevreforjan2025saturnssuperjupitersoccurfrequently}.}

Some formation and evolution scenarios have disfavored these types of planetary system architectures. \revise{Many simulations of} giant planets \revise{have demonstrated difficulties with forming inner super-Earths by limiting} the flow of pebbles into the inner disks\footnote{\reviseTwo{Some recent simulations have suggested that, if the cores of outer giant planets formed later in the gas disk phase, there would be ample time for multiple generations of super-Earths to form before pebble accretion would largely cease; however, the significant amount of gas in the system would cause many of these planets to be accreted onto the star and thus not survive long enough to undergo late-stage planet formation \citep{VoelkelKlahr2022_A&A}.  \reviseThree{This process could leave the inner disk depleted of solids, leading to the formation of much less massive inner planets than otherwise expected.}}} (\revise{\citealt{BitschMorbidelli2018_A&A, AtaieeBaruteau2018_A&A};} \citealt{MuldersDrazkowska2021_ApJL})\revise{; acting} as barriers to super-Earths migrating into the inner region \citep{IzidoroRaymond2015_ApJL}\revise{; and reducing the multiplicity of inner super-Earths through dynamical interactions \citep{BitschTrifonov2020_A&A, Bitsch2023}.}  However, given the observed correlations, either these barriers are surmountable or the true formation scenarios are significantly different from what we expect (e.g., \citealt{Best2024}).%\revise{More recent simulations \citep[e.g.,][]{Bitsch2023} have shown that these barriers are either surmountable or }  

Although giant planets do not prevent the formation of inner planets, they can influence their orbital characteristics. Dynamical excitation of inner super-Earths by outer giant planets \citep[e.g.,][]{JohansenDavies2012_ApJ, MustillDavies2017_MNRAS, Huang2017, PuLai2018_MNRAS,Poon&Nelson2020, AnXie2023_AJ, Bitsch2023} could help to explain the observed excess of single transiting planets observed in the \textit{Kepler} exoplanets (e.g., \citealt{HansenMurray2013_ApJ,BallardJohnson2016_ApJ}) and the higher observed eccentricities of these single transiting planets \citep[e.g.,][]{XieDong2016_PNAS, VanEylen2019}. \revise{However}, not all configurations lead to orbital excitation. \citet{MasudaWinn2020_AJ} find observational evidence for higher mutual inclinations of outer Jupiters that accompany single transiting systems vs. those that accompany multi-transiting systems.  \citet{ChildsQuintana2019_MNRAS} simulate the effects on terrestrial planet formation of giant planets with varying masses at the location of Jupiter and Saturn. They find that more massive giant planets produce inner terrestrial planets with closer-in and more circular orbits. Moreover, inner super-Earths can resist perturbations from outer giant planets if they are in mean motion resonance, which can stabilize their orbits \citep{RodetLai2021_MNRAS}. Furthermore, other properties---like the amount of gas \citep{DawsonLeeChiang2016} and solids \citep{Moriarty&Ballard2016,MacDonald2020} available during the giant impact stage of planet formation---can influence the degree of orbital excitation, independent of giant planets.

In this study, we aim to advance the understanding of if and how outer giant planets contribute to the observed orbital excitation of inner super-Earths. We consider a scenario in which the outer giant planets \reviseTwo{completed formation} during the gas disk stage while the inner super-Earths are still isolation mass embryos. The inner super-Earths complete their formation later, via giant impacts of the embryos, after the gas disk has fully, or at least partially (e.g., \citealt{DawsonLeeChiang2016,Lee&Chiang2016}), dissipated \reviseTwo{(see Section \ref{subsec:theoreticalFramework} for a more detailed description of our assumed formation scenario)}. This scenario leaves a period of time at the end of the gas phase and into the post-gas phase of the protoplanetary disk during which the growing embryos in the inner system are gravitationally influenced by outer giant planets before reaching their final orbital and physical properties.  \reviseTwo{Our study focuses on this period of time: the tail end of the gas phase and the subsequent post-gas evolution that we refer to as ``late-stage planet formation''.}

This scenario differs from certain prior studies where giant planets were introduced into mature, observed planetary systems. Our giant planets are introduced \reviseTwo{during the last stage of planet formation while the inner systems are still forming. By allowing the inner planets to form in the presence of outer giant planets, we ensure the completeness of our final underlying planetary systems and do not assume that the systems mirror the observed multi-transiting planets,} where we are likely missing planets and have poor constraints on orbital properties like eccentricity. Like \citet{Bitsch2023}, we introduce giant planets during the formation stage, though here we focus on in situ formation rather than migration.
%This scenario differs from certain prior studies where giant planets were introduced into mature, observed planetary systems. Our giant planets are introduced \revise{during late-stage planet formation ensuring the completeness of our final underlying planetary systems}, rather than systems containing only the observed multi-transiting planets, where we are likely missing planets and have poor constraints on orbital properties like eccentricity. Like \citet{Bitsch2023}, we introduce giant planets during the formation stage, though here we focus on in situ formation rather than migration. 
In the presence of a gas disk that damps free eccentricities and inclinations, small planets may be able to adjust to giant planet perturbations that would be highly disruptive if the giant planet is suddenly introduced after the formation stage. 

Alternatively, we may find that giant planets can be highly disruptive even when introduced during the formation stage. We compare directly to the simulations from \citet{MacDonald2020}, which have the same initial conditions except no outer giant planets, allowing us to directly assess the effects of giant planets. \revise{Using $N$-body simulations,} we \revise{investigate} two configurations of giant planets: one based on \citet{Huang2017}'s study of excitation of inner super-Earths by dynamically active giant planets and one based on the Solar System giant planets which have lower mutual inclinations and may have a gentler effect (e.g., \citealt{ChildsQuintana2019_MNRAS,MasudaWinn2020_AJ}).  We forward model selection effects and account for the relatively low occurrence rate of giant planets when comparing to observed planetary systems.

In Section \ref{sec:methods}, we describe the theory and assumptions we use when modeling late-stage planet formation, the simulations we perform, and our procedure for forward modeling detections of transiting planets from our simulated systems. In Section \ref{sec:DAGresults}, we present the results from our simulations that include dynamically active giant planets. In Section \ref{sec:ssg}, we discuss results from simulations with Solar System giant planets.  In Section \ref{sec:comparison}, we directly compare the results from our different giant planet configurations.  In Section \ref{sec:mix}, we explore mixing together systems with and without giant planets in an attempt to produce distributions that match observations. Finally, we present our summary and conclusions in Section \ref{sec:prelimConclusionsandFutureWork}.

\section{Methods \label{sec:methods}}

We use $N$-body simulations to model the late stage in situ formation of super-Earths in a protoplanetary disk with outer giant planets.  We perform sets of simulations using two different configurations of outer giant planets: one that includes the four Solar System giants---which we call Solar System Giants or SSG---and one that includes three giant planets with random masses, orbital periods, and orbital elements following \citet{Huang2017}---which we call Dynamically Active Giants or DAG.

\subsection{Theoretical Framework \label{subsec:theoreticalFramework}}

During the gas phase of the protoplanetary disk, planetesimals undergo oligarchic growth, forming planetary embryos.  The growth rate and final mass of the embryos is determined by the amount of solids in disk, parameterized by the disk's solid surface density $\Sigma_z$ \citep{KokuboIda2002_ApJ}. We adopt a power-law model of $\Sigma_z$,

\begin{equation} \label{eq:solidSurfaceDensity}
    \Sigma_z = \Sigma_{z,1} \qty( \frac{a}{\text{AU}} )^\alpha, 
\end{equation}
where $a$ is the semimajor axis, $\Sigma_{z,1}$ is the solid surface density at 1 au, and $\alpha$ is a constant. 

After reaching their isolation mass, embryos can grow via giant impacts. If $\Sigma_z$ is large enough (e.g., \citealt{MacDonald2020}) and/or if the gas surface density is low enough for an extended period, the embryos can grow via giant impacts in the presence of a partially depleted gas disk.  The gas damps the growing planet's eccentricity and inclination, leading to a final system configuration which is dynamically cold (low eccentricities, low mutual inclinations, compact spacings) that persists long after the dissipation of the gas disk (e.g., \citealt{DawsonLeeChiang2016}). However, if the initial solid surface density of the protoplanetary disk is low relative to the gas \citep{MacDonald2020}---or the low amount of solids in the disk form smaller isolation-mass embryos which leads to long merger timescales (e.g., \citealt{Dawson2015})---the planetary embryos will continue to merge after the gas disk has dissipated.  In this case, the resulting planets will tend to be more dynamically hot (larger eccentricities, mutual inclinations, and spacings). 

\reviseTwo{In general, protoplanetary gas disks dissipate gradually until the gas becomes significantly depleted, at which time the gas disappears rapidly and completely} \citep{Owen2012}.  In our simulations, we \reviseTwo{approximate this process by modeling the last stage of the gas disk with a step function.  We first assume a constant depletion factor $d$ for the gas surface density profile.  Then, at} $t=\tau_\text{dis}$, \reviseTwo{the gas surface density} $\Sigma_g$ becomes zero.  We model the gas surface density $\Sigma_g$ as:

\begin{equation} \label{eq:gasSurfaceDensity}
    \Sigma_g =
    \begin{dcases}
        1700\  d^{-1}\, \qty(\frac{a}{\mathrm{AU}})^{-3/2}\, \mathrm{g\, cm}^{-2} & t < \tau_\text{dis}\\
        0 & t > \tau_\text{dis},
    \end{dcases} 
\end{equation}
where $d$ is the gas depletion factor, $d=1$ corresponds to the full minimum mass solar nebula (MMSN), and $d>1$ to a depleted nebula. Following \citet{DawsonLeeChiang2016,MacDonald2020}, we use  $\tau_\text{dis}=1 \text{\,Myr}$ for the depleted gas disk stage.  \reviseTwo{While $t<\tau_\text{dis}$}, the gas surface density is depleted enough such that migration, and thus its effect on the final planet properties, is negligible \citep{MacDonald2020}.

In order to compare our simulations to \citet{MacDonald2020}'s simulations without giant planets, we adopt identical parameters and initial embryos to that study:

\begin{equation*}
    \begin{gathered}
        14\, \text{g cm}^{-2} \leq \Sigma_{z,1} \leq 284\, \text{g cm}^{-2}, \\
        \alpha = -1.5, \\ 
        d = 100.
    \end{gathered}
\end{equation*}

\citet{MacDonald2020} found that simulations with $\alpha = -1.5$, $d=100$, and solid surface densities drawn from the range above produced mock observed planets with properties similar to the {\it Kepler} sample. We also perform a suite of simulations for one of the giant planet configurations that does not include a gas damping stage (see Section \ref{subsec:simulations}).

\subsection{Simulations \label{subsec:simulations}}

We perform $N$-body simulations using \texttt{REBOUND}'s hybrid symplectic integrator \texttt{MERCURIUS}  \citep{rebound:paper2012, rebound:paper2015a, rebound:paper2015b, rebound:paper2019}. \texttt{MERCURIUS} uses \texttt{WHFAST}, a symplectic integrator, and switches to using \texttt{IAS15}, a high order non-symplectic integrator with an adaptive time step, when two particles experience a close encounter.  \revise{In order to readily compare our results with \citet{MacDonald2020}, we adopt the same initializations for our simulations. Specifically,} we use a time step of 0.5 days for \texttt{MERCURIUS} and accuracy parameter of $10^{-9}$ for \texttt{IAS15}. We integrate our simulations for 1 Myr in the presence of the gas disk\footnote{Gas damping implementation is available as \texttt{gas\_damping\_timescale} within the \texttt{REBOUNDx} package at \url{https://github.com/dtamayo/reboundx}.} (implementation of gas damping is described in detail in Appendix \ref{app:gasDamp}) and integrate an additional 27.4~Myr after the gas has dissipated.  \revise{This integration time is \reviseTwo{identical to} previous works\footnote{\reviseTwo{Specifically, \citealt{MacDonald2020} integrated for $10^{10}$~days which corresponds to roughly 27.4~Myr.}} and integrating for longer has been found to negligibly change the resulting underlying and observed distributions \citep{DawsonLeeChiang2016, MacDonald2020}.} During collisions, we assume perfect accretion without fragmentation and a collisional radius corresponding to a density $\rho = 1 \text{ g cm$^{-3}$}$.\footnote{\reviseTwo{This density was the default value used by the $N$-body integrator \texttt{mercury6} \citep{Chambers_Mercury6}, and thus adopted by prior works that used \texttt{mercury6}.  Although we use a different $N$-body integrator, we adopt the identical value for consistency.}} % The choice of this value ultimately only affects the computation time of the chosen integrator by lowering the time necessary for a collision to occur as compared to using a larger density more consistent with rocky bodies, and does not impact the final results of the simulations.}}

For each configuration, we perform 290 simulations, each with a different value of $\Sigma_{z,1}$ drawn from a log uniform distribution ranging from $14$--$284$ g cm$^{-2}$.  We perform 50 additional simulations for intermediate solid surface density values between $\sim$55--103 g cm$^{-2}$.  \revise{These} $\Sigma_{z,1}$ values \revise{are identical to those used in \citet{MacDonald2020}.  This range of values was also found to produce} more observable systems \citep{MacDonald2020}. These intermediate values of $\Sigma_{z,1}$ also correspond approximately to the average solid surface density inferred from observed \textit{Kepler} systems \citep{HeFord2022_AJ}.  

\revise{After the simulations have run for each configuration, we find the minimum value of $\Sigma_{z,1}$ that produces any unphysically large planet ($>20 M_\oplus$) and exclude all simulations whose solid surface density normalization factor is greater than this minimum value.} \reviseTwo{Ultimately, we do not include any simulations with $\Sigma_{z,1}>170$ g cm$^{-2}$.  Thus, the full range of $\Sigma_{z,1}$ values explored aligns with those produced in simulations in other works \citep[e.g.,][]{GibbonsMamatsashvili2015_MNRAS, HeFord2022_AJ}.}

\subsubsection{Planetary Embryos \label{subsubsec:planetaryEmbryos}}

For our simulations of late stage planet formation, we use the \revise{identical} initial \revise{physical and orbital properties as the} planetary embryos \revise{from} \citet{MacDonald2020}.  The semimajor axis of the innermost planetary embryo $a_\mathrm{inner}$ is randomly selected from a uniform distribution  with a range of 0.04--0.06 au.  The initial spacing of the embryos, in units of mutual Hill radii $\Delta$, is $\Delta_0 = 3$ where
\begin{equation} \label{eq:mutualHillSpacing}
    \Delta \equiv \frac{a_2 - a_1}{R_H},
\end{equation}
\noindent where $a_1$ and $a_2$ are the semimajor axes of inner and outer adjacent bodies, respectively, $R_H$ is the mutual Hill radius
\begin{equation} \label{eq:HillRadius}
    R_H = \frac{a_1 + a_2}{2} \qty(\frac{M_{p,1} + M_{p,2}}{3 M_\star} )^{1/3},
\end{equation}
\noindent $M_p$ is the mass of the planet, and $M_\star$ is the mass of the central star.

The number of initial embryos $\Nemb$ ranges from $\Nemb = 72$ to $\Nemb = 348$, distributed between $a_\mathrm{inner}$ and 1 AU.  Assuming a stellar mass of $1 M_\odot$, the mass of each embryo \revise{is} calculated to be

\begin{equation}
    M_{\text{emb}} = 0.16M_\oplus \qty(\frac{\Sigma_{z,1}}{33 \text{ g cm$^{-2}$}})^{3/2} \qty(\frac{a}{\text{AU}})^{3/4}.
\end{equation}

As a point of reference, the typical embryo in our simulations has a mass of $\sim 0.5 \,M_\text{Mars}$.

\revise{Each embryo has an initial eccentricity of $e_0=0$ and is assigned an initial inclination $i_0 = 0.01h/\sqrt{3}$, where  }

\begin{equation}
    h = \left(\frac{M_{p,1} + M_{p,2}}{3M_\star}\right)^{1/3}
\end{equation}

\revise{The initial spacings, eccentricities, and inclinations of the embryos approximate the properties of the embryos that would have been produced had we fully modeled \reviseTwo{the entirety of the evolution of the gas disk, and not just the final 1~Myr} \citep{DawsonLeeChiang2016, MacDonald2020}.}

% dawson 2015 for why those initial inclinations
% Give short summary of this section and say again that these are the same embryos as macdonald 2020

\subsubsection{Outer Giant Planets \label{subsubsec:giantPlanets}}

\begin{deluxetable*}{cccccccc}
\tablecolumns{8}
\tablecaption{\label{tab:GiantPlanetParam} Top: Masses and orbital parameters of the giant planets used in the SSG simulations. Bottom: The distributions used to generate the masses and orbital properties of the giant planets used in the DAG simulations.}
\tablehead{\multicolumn{8}{c}{Solar System Giants} \\
\colhead{Planet} & \colhead{$m_p$ ($M_\odot$)} & \colhead{$a$ (AU)} & \colhead{$e$} & \colhead{$i$\tablenotemark{\footnotesize a} (deg)} & \colhead{$\omega_p$\tablenotemark{\footnotesize b} (deg)} & \colhead{$\Omega$\tablenotemark{\footnotesize c} (deg)} & \colhead{$M$\tablenotemark{\footnotesize d} (deg)}}
\startdata
Jupiter     &   $9.54 \times 10^{-4}$   &   5.2       &     0.048   &   1.32     &  14.75   &    100.56     &   -80.91  \\
Saturn      &   $2.85 \times 10^{-4}$   &   9.54      &     0.054   &   2.46     &  92.43   &    113.72     &   -156.2  \\
Uranus      &   $4.35 \times 10^{-5}$   &   19.19     &     0.047   &   0.74     &  170.96  &    74.23      &   68.04   \\
Neptune     &   $5.15 \times 10^{-5}$   &   30.07     &     0.009   &   1.78     &  44.97   &    131.72     &   128.19  \\
\hline
\hline
\noalign{\vskip 0.5mm} 
\multicolumn{8}{c}{Dynamically Active Giants} \\
\noalign{\vskip 1mm} 
Number of Planets         &   $m_p$ ($M_\text{Jup}$)   &   $a$\tablenotemark{\footnotesize e} (AU)     &     $e$   &   $i$ (deg)     &  $\omega_p$ (deg)    &    $\Omega$ (deg)      &   $M$ (deg)   \\
\noalign{\vskip 0.75mm} 
\hline
3         &   $\sim U(0.3, 3.0)$   &   $\sim U(2, 5)$    &     $\sim$\text{Rayleigh}$(\sigma = 0.01)$   &   $\sim$\text{Rayleigh}$(\sigma = 0.01)$     &  $\sim U(0, 360)$  &    $\sim U(0, 360)$     &   $\sim U(0, 360)$  \\
\hline
\hline
\enddata
\tablenotetext{a}{inclination of planet with respect to the invariable plane of the Solar System}
\tablenotetext{b}{argument of periapsis of the planet}
\tablenotetext{c}{longitude of ascending node of planet}
\tablenotetext{d}{initial mean anomaly of the planet}
\tablenotetext{e}{rejected any draws that would produce $\Delta < 3$}
\end{deluxetable*}  

We assume that the giant planets have fully formed and reached their current positions during the gas-dominant phase of the protoplanetary disk \citep{LissauerStevenson2007_prpl}. We investigate the impact of two configurations of giant planets.  We refer to these two cases as: the Solar System giant planets (SSG) and the dynamically active giant planets (DAG).  The Solar System giant planets are widely separated from one another and from the inner system planetary embryos.  The masses and orbital parameters of the Solar System giant planets are listed in Table \ref{tab:GiantPlanetParam}.

For the dynamically active configuration, we follow the procedure outlined in Section 2.2.2.\ in \citet{Huang2017}, who  performed \textit{N}-body simulations to examine how unstable outer giant planets affect the properties of fully-formed close-in super-Earths.  We \revise{generate} three giant planets with: masses between 0.3 $M_{\text{Jup}}$ and 3 $M_{\text{Jup}}$ from a uniform distribution; semimajor axes between 2 and 5 AU from a uniform distribution (rejecting any draws with mutual Hill radii $<3$); eccentricities and inclinations drawn from a Rayleigh distribution with a width $\sigma = 0.01$; and mean anomalies, arguments of periapse, and longitudes of ascending node each from a random uniform distribution between 0 to 360$^\circ$. \revise{\citet{Huang2017} chose these distributions of initial orbital and physical properties such that the final properties of the outer giant planets broadly match the population of giant planets observed with radial velocity surveys.} 

To summarize, we run a total of three suites of simulations to compare to each other and to those from \citet{MacDonald2020}, which we term No Giant Plants (NGP) simulations: (1) simulations with three Dynamically Active Giant planets and a gas damping phase (DAG); (2) simulations with three Dynamically Active Giant planets without a gas damping phase (DAG*); and (3) simulations with Solar System Giant planets and a gas damping phase (SSG). Each suite consists of 290 simulations, for a total of 870 simulations.

\subsection{Forward Modeling Detection of Transiting Planets \label{subsec:forwardModeling}}

We compare our simulated planets to the planet candidates in the DR25 \textit{Kepler} catalog \citep{Thompson2018_KeplerDR25}.  We use the following criteria\revise{, chosen to be consistent with \citet{MacDonald2020},} to select for \textit{Kepler} systems that orbit bright main-sequence FGK stars and that host at least one super-Earth:
\begin{enumerate}
    \item $4100\, \mathrm{K} < T_\mathrm{eff} < 6100 \, \mathrm{K}$
    \item $\log g > 4$
    \item $K_\mathrm{p} < 15$
    \item $R_p < 4 R_\oplus$, for one or more planets
\end{enumerate}
where $T_\mathrm{eff}$ is the stellar effective temperature, $\log g$ is the stellar surface gravity, $K_\mathrm{p}$ is the \textit{Kepler} magnitude, and $R_p$ is the planet radius.

In order to compare our simulated planets to this subset of the \textit{Kepler} sample, we forward model the detection of transiting planets following \citet{MacDonald2020}.  For each simulated planetary system, we generate $10^4$ instances of the system randomly oriented in space with respect to the observer.  We calculate the impact parameter 
\begin{equation}
    b = \frac{a}{R_\star} \cos{i} \qty(\frac{1-e^2}{1+e\sin{\omega_p}}),
\end{equation}
where $R_\star$ is the stellar radius, $i$ is the inclination of the system with respect to the sky plane, $e$ is the eccentricity, and $\omega_p$ is the argument of periapsis of the planet.  If $b<1$, the planet transits.

For each instance, we determine the detection probability of each planet in the system following \citet{2015ApJ...809....8B} and using the injection-recovery results of \citet{2015ApJ...810...95C,2016ApJ...828...99C}.  The detection probability is a function of impact parameter, planet radius\footnote{We calculate a planetary radius using the mass-radius relationship from \citealt{2017ApJ...834...17C}.}, 
orbital period, and combined differential photometric precision.  We marginalize the detection probability over all {\it Kepler} target stars meeting the criteria at the beginning of this subsection (approximately 60k targets).  We draw a number randomly from a uniform distribution between 0 and 1. We ``detect'' all transiting planets in that mock observed instance with detection probability greater than that number.

The input values of $\Sigma_{z,1}$ in our simulations are drawn from a log uniform distribution, which may not be representative of the true distribution of solid surface densities in observed systems. We use a Gaussian weighting factor $W$ centered at ${\Sigma^-_{z,1}}$, the mean of the true underlying distribution of solid surface densities:
\begin{equation} \label{eq:weightingFactor}
    W = \exp{\qty[-\frac{(\Sigma_{z,1} - \Sigma^-_{z,1})^2}{2 \sigma^2_{\Sigma_{z,1}}} ]}
\end{equation}

\noindent where $\sigma_{\Sigma_{z,1}}$ is the width of the Gaussian. We adopt the values ${\Sigma^-_{z,1}} = 0$ and $ \sigma_{\Sigma_{z,1}} = 65 \text{ g cm$^{-2}$}$ from \citet{MacDonald2020} that provided a qualitatively good match for the \textit{Kepler} observables.  For each instance, we generate a random variable $x$ drawn from a uniform distribution between 0 and 1. If $x < W$, that detection is included.

We compare our results from the forward-modeled transit detections to the \textit{Kepler} catalog observables. We choose four derived parameters to include in this comparison: the period ratio of adjacent planets, the mutual Hill spacing between adjacent planets, the ratio between the transit durations of adjacent planets, and the transit multiplicity. The period ratio is calculated using a directly observable property of transiting planets and can indicate if two planets are in or near resonance. For observed planets without a published mass, we use the mass-radius relationship from \citet{2017ApJ...834...17C} to compute a mass to use to determine the mutual Hill spacing.

The total transit duration for an eccentric orbit can be approximated by:

\begin{equation} \label{eq:fullTransitDuration}
    \begin{split}
    T_\text{dur} & \approx \frac{P R_\star}{\pi a} \sqrt{\qty(1 + \frac{R_p}{R_\star})^2 - \qty(\frac{a}{R_\star} \cos{i} )^2} \\
    & \times \frac{\sqrt{1-e^2}}{1+e \sin{\omega_p}}, 
    \end{split}
\end{equation}

\noindent where $P$ is the period, $R_\star$ and $R_p$ are the stellar and planetary radii, respectively, $i$ is the inclination of the planet with respect to the plane perpendicular to the line of sight, $e$ is the eccentricity of the planet, and $\omega_p$ is the argument of periapsis of the planet \citep{Winn2010_exop}.

The normalized ratio of transit duration between adjacent objects is 
\begin{equation} \label{eq:ratioTransitDuration}
    \xi = \frac{T_{\text{dur},1}}{T_{\text{dur},2}} \qty(\frac{P_1}{P_2})^{-1/3},
\end{equation}

\noindent where the subscripts 1 and 2 indicate the inner and outer adjacent planets, respectively. Because the transit duration depends on the impact parameter and eccentricity, the distribution of $\xi$ among observed planets is affected by planets' mutual inclinations and eccentricities. Here we use $\log_{10}{\xi}$. In general, the distribution of $\log_{10}{\xi}$ peaks close to zero and is asymmetric, with more coplanar adjacent planets shifting the distribution of $\log_{10}{\xi} \geq 0$. Larger eccentricities, larger mutual inclinations, and observational uncertainties widen the distribution. The last parameter we compare is the transit multiplicity which is set by a system's underlying planet multiplicity, the mutual inclinations between planets, and the inclination relative to the observer.

\section{Dynamically Active Giants} \label{sec:DAGresults}

\begin{figure*}[htbp]
\epsscale{0.75}
\plotone{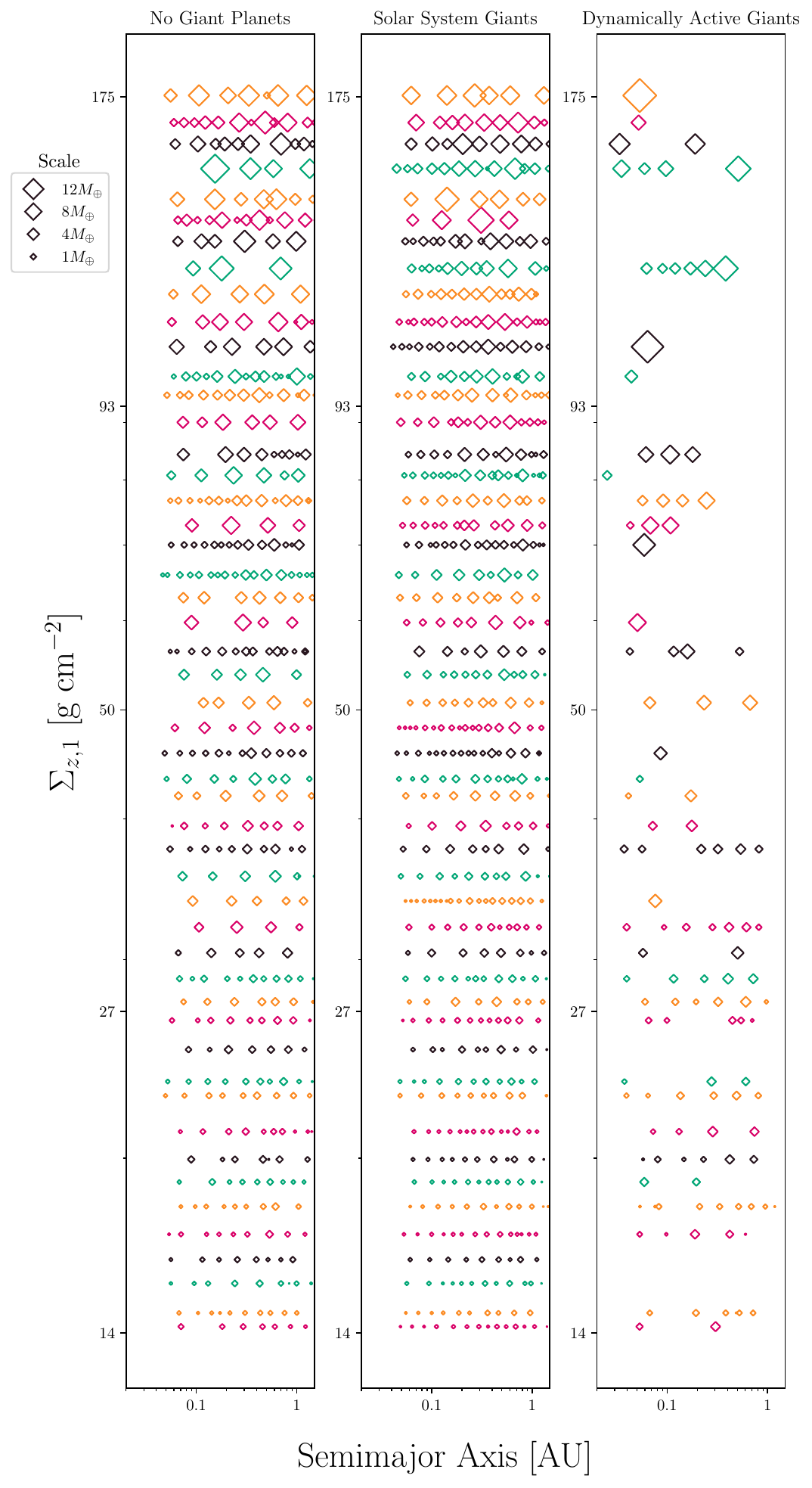} 
\caption{Final configurations of 50 planetary systems, randomly selected from our sample of simulated systems and sorted by the solid surface density normalization factor $\Sigma_{z,1}$ for the NGP simulations from \citet{MacDonald2020} (left), the SSG simulations (center), and the DAG simulations (right).  The positions of the surviving non-giant planets are plotted as a function of semimajor axis; the size of each point is proportional to the planet's mass (scale in legend).  Each row of points corresponds to a set of simulations with identical initial embryos.  Some rows in the right panel are empty because there are no surviving inner planets ($\Nsp = 0$).
\label{fig:underlying}}
\end{figure*}

The types of systems produced in our three sets of systems (NGP, DAG, and SSG) are depicted in Figure \ref{fig:underlying} and summarized in Table \ref{tab:properties}. In this section, we focus on the results from the DAG simulations. The DAG simulations tend to form systems with fewer, more massive, more widely spaced inner planets than the NGP case (Figure \ref{fig:underlying}).

\begin{deluxetable}{cccc}[htbp]
\tablecolumns{4}
\tablecaption{\label{tab:properties} Properties of underlying planets and the forward-modeled observed planets averaged over each set of simulations. Except where otherwise specified, the reported value is the median, and uncertainties denote the 68\% high density interval computed with \rm{ArviZ} \citep{arviz:paper, arviz:zenodo}.}
\tablehead{
\colhead{} &  \colhead{NGP} & \colhead{DAG} & \colhead{SSG}}
\startdata
\sidehead{Underlying Planets}
$\Nsp$\tablenotemark{\footnotesize a} & $7^{+1}_{-4}$ & $1^{+2}_{-1}$ & $10^{+2}_{-4}$\\[0.1cm]
$a$ (AU) & $0.3^{+0.3}_{-0.3}$ & $0.2^{+0.2}_{-0.2}$ & $0.3^{+0.3}_{-0.3}$ \\[0.1cm]
$m_p$ ($M_\oplus$) & $1.5^{+1.0}_{-1.5}$ & $1.6^{+1.4}_{-1.5}$ & $1.4^{+0.9}_{-1.2}$ \\[0.1cm]
$i_\mathrm{mut}$ (deg) & $2^{+2}_{-2}$ & $1.8^{+1.6}_{-1.9}$ & $0.4^{+1.2}_{-0.4}$\\[0.1cm]
Period Ratio & $1.7^{+0.4}_{-0.5}$ & $2.1^{+0.7}_{-0.8}$ & $1.5^{+0.2}_{-0.3}$ \\[0.1cm]
$\Delta$\tablenotemark{\footnotesize b} & $22^{+8}_{-11}$ & $32^{+18}_{-14}$ & $18^{+6}_{-7}$\\[0.1cm]
\sidehead{Mock Observed Planets}
$a$ (AU) & $0.09^{+0.03}_{-0.04}$ & $0.06^{+0.02}_{-0.02}$ &  $0.07^{+0.03}_{-0.03}$\\[0.1cm]
$m_p$ ($M_\oplus$) & $2.3^{+1.1}_{-1.8}$ & $2.5^{+1.3}_{-2.0}$ &  $1.7^{+0.8}_{-1.2}$\\[0.1cm]
$i_\mathrm{mut}$ (deg) & $1.3^{+1.8}_{-1.3}$ & $1.2^{+0.9}_{-1.2}$ & $0.3^{+1.3}_{-0.3}$\\[0.1cm]
Period Ratio & $2.1^{+0.7}_{-0.9}$ & $3.6^{+1.5}_{-2.0}$ & $1.7^{+0.4}_{-0.5}$ \\[0.1cm]
$\Delta$\tablenotemark{\footnotesize b} & $28^{+12}_{-13}$ & $45^{+11}_{-25}$ & $23^{+8}_{-10}$\\[0.1cm]
$\log_{10}\xi$\tablenotemark{\footnotesize c} & $0.02^{+0.12}_{-0.10}$ & $0.03^{+0.13}_{-0.11}$ & $0.02^{+0.10}_{-0.09}$\\[0.1cm]
Transit Multiplicity \tablenotemark{\footnotesize d} & $1.5 \pm 1.0$ & $1.2 \pm 0.5$ & $1.9 \pm 1.5$\\[0.1cm]
$e_\mathrm{single}$\tablenotemark{\footnotesize e} & $0.08^{+0.04}_{-0.08}$ &$0.11^{+0.06}_{-0.11}$ & $0.07^{+0.02}_{-0.07}$ \\[0.1cm]
$e_\mathrm{multi}$\tablenotemark{\footnotesize e} & $0.04^{+0.03}_{-0.03}$  & 
$0.07^{+0.04}_{-0.07}$ & $0.01^{+0.03}_{-0.01}$ \\[0.1cm]
\enddata
\tablenotetext{a}{$\Nsp \equiv$ Number of non-giant planets that form and survive}
\tablenotetext{b}{$\Delta \equiv$ Mutual Hill Spacing (Eq.\ \ref{eq:mutualHillSpacing})}
\tablenotetext{c}{$\xi \equiv$ Normalized transit duration ratio (Eq.\ \ref{eq:ratioTransitDuration})}
\tablenotetext{d}{Mean and standard deviation}
\tablenotetext{e}{Eccentricities averaged over single transit and multi-transit systems }
\end{deluxetable} 

\subsection{Effects on underlying systems}

\begin{figure*}[htbp]
 \epsscale{1.2}
%\plotone{figures/mass-vs-semimajoraxis_evolution_DAG_NGP.pdf}
%\plotone{figures/ecc-vs-semimajoraxis_evolution_DAG_NGP.pdf}
\plotone{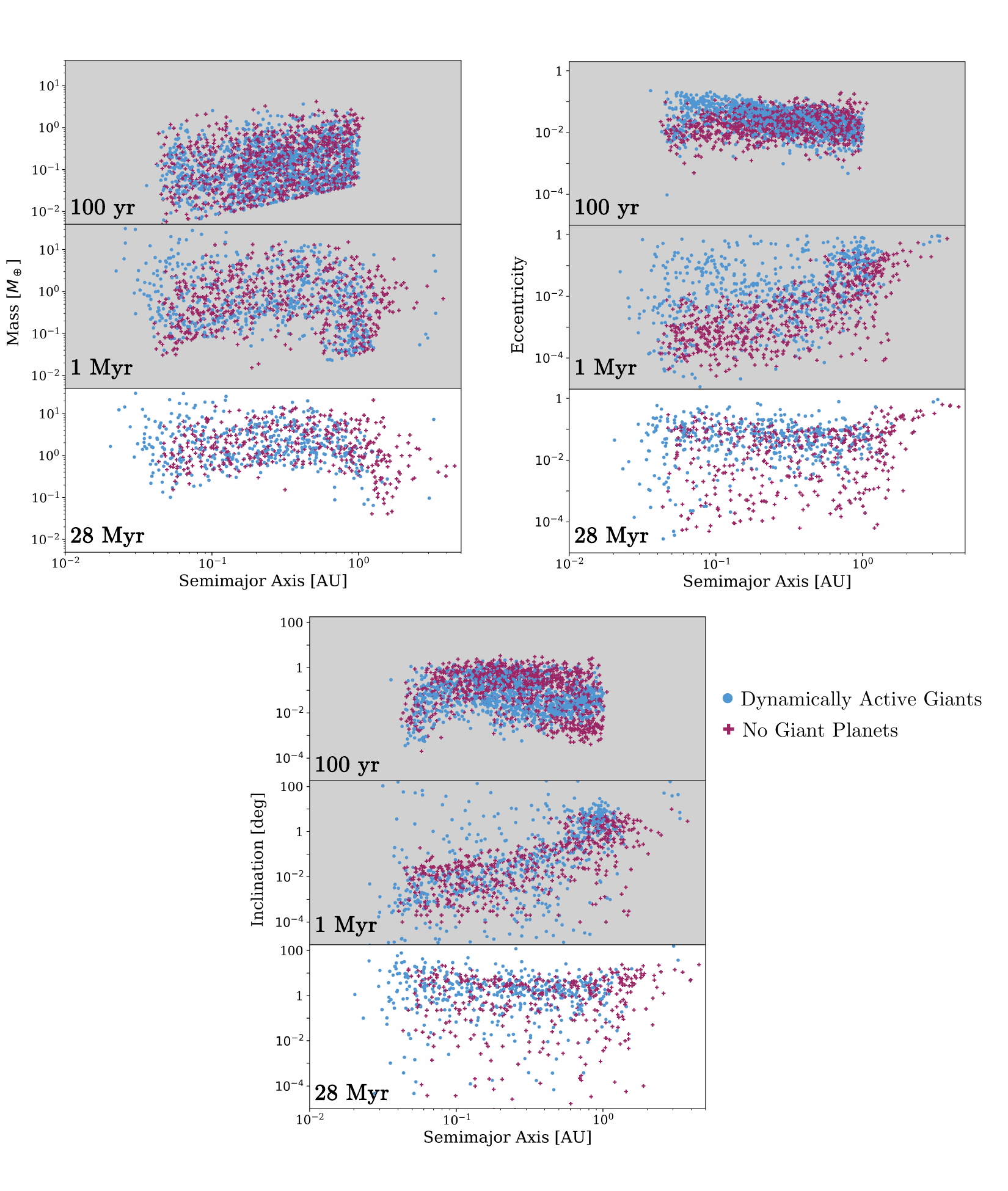}
\caption{Time evolution of mass (top left), eccentricity (top right), and inclination (bottom) vs.\ semimajor axis in the NGP (dark pink pluses) and the DAG (blue points) simulations. The filled-in gray background of some of the plots indicates the snapshot was taken during the gas phase.  \revise{The inclinations were measured with respect to the reference plane in \texttt{REBOUND}, which corresponds to the $x$-$y$ plane.} \label{fig:time}}
\end{figure*}

Figure \ref{fig:time} shows the time evolution of mass vs.\ semimajor axis, eccentricity vs.\ semimajor axis, and inclination vs.\ semimajor axis of planetary embryos as embryos merge to become full-fledged planets. Compared to the simulations without any outer giant planets, DAG simulations produce more planetary embryos with excited eccentricities and inclinations early on ($\sim100$ years) during the gas phase. Despite gas damping, planets in the DAG simulations maintain higher eccentricities and inclinations than those in the NGP simulations. These higher eccentricities and inclinations persist after the gas stage.

The DAG simulations produce a population of short period planets ($a<0.05$ AU, $P<4$ days) within the initial disk truncation region.  In contrast, the planets grown in the NGP case do not intrude into this initially empty region. About halfway through the \revise{DAG} gas phase ($\sim0.5$ Myr), embryos begin to be scattered interior to 0.05 AU. The scatterings continue throughout the rest of the gas phase and the post-gas phase.  

\begin{figure}[htbp]
\plotone{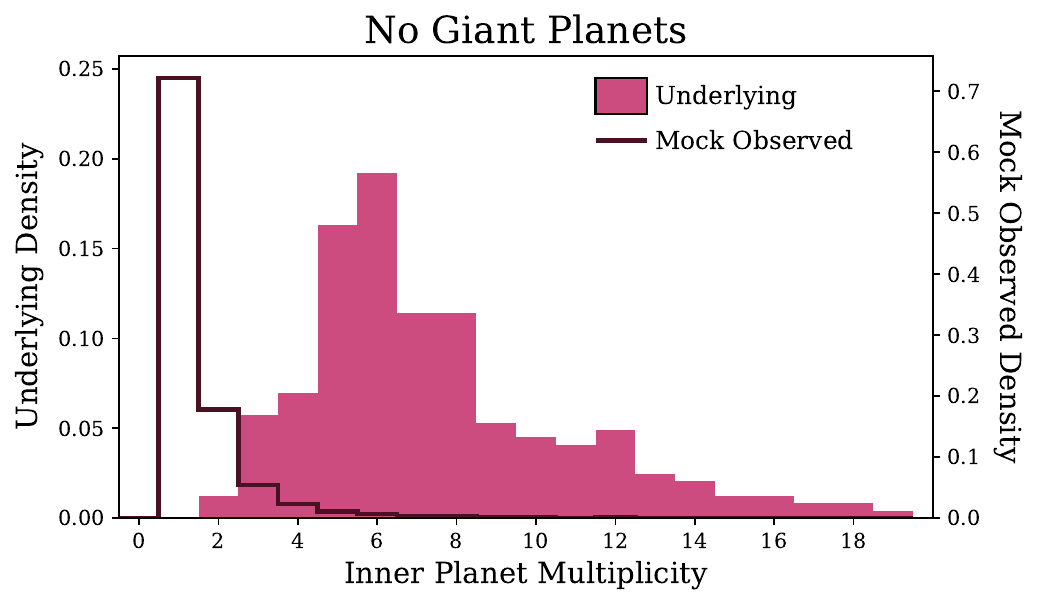}
\plotone{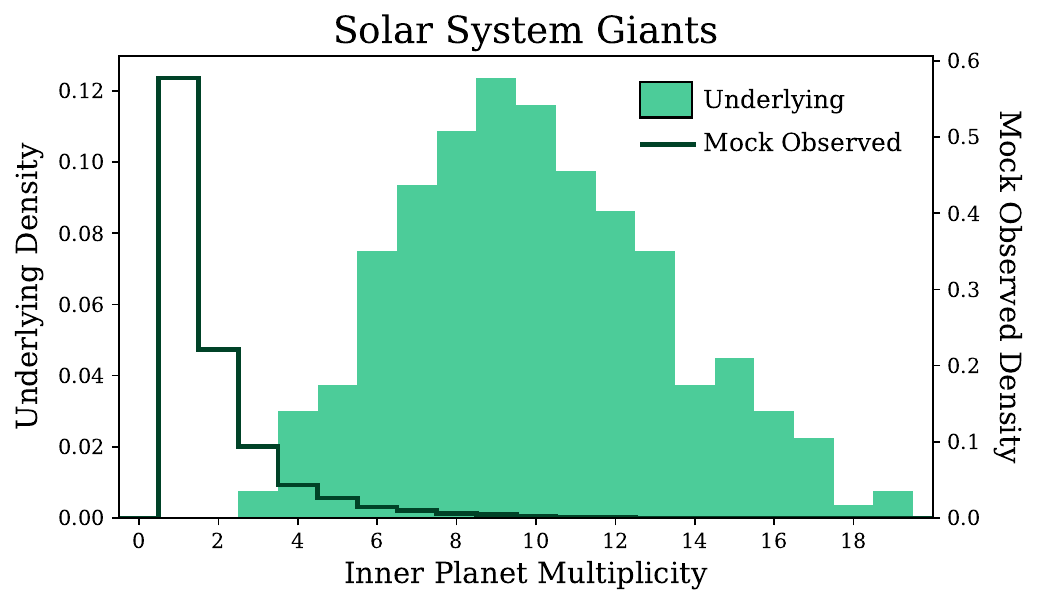}
\plotone{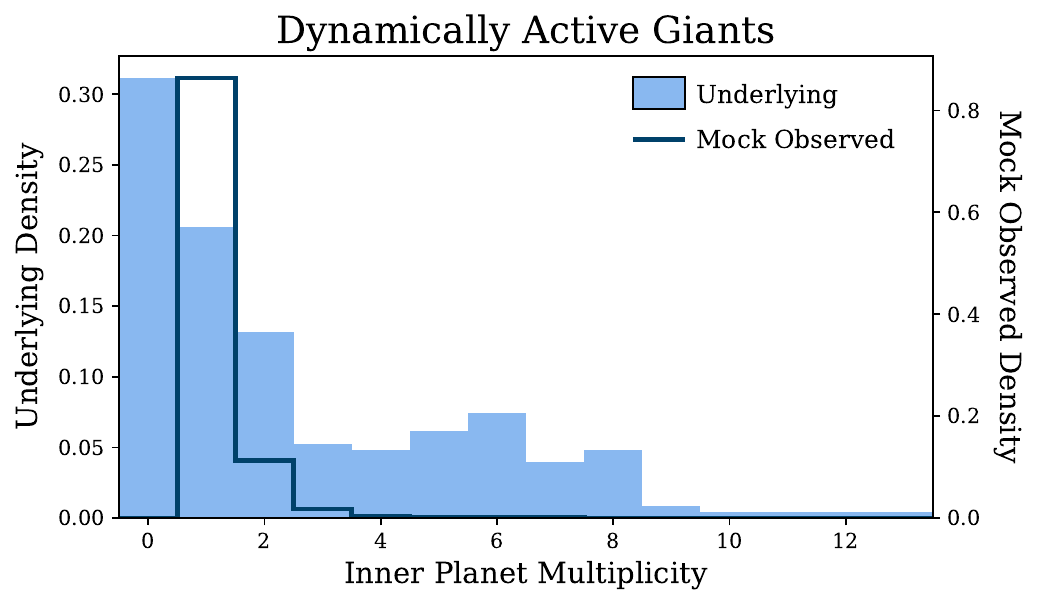}
\caption{Inner planet multiplicity for each case (NGP: top, SSG: middle, DAG: bottom).  The filled in histograms and left-handed axes correspond to the underlying inner planet multiplicity.  The unfilled histograms and right-handed axes correspond to the mock observed inner planet multiplicity.  The SSG systems formed the largest number of inner planets, while the DAG systems formed by far the fewest. \label{fig:NumSurvivingPlanets}}
\end{figure}

\begin{figure*}[htbp]
\plotone{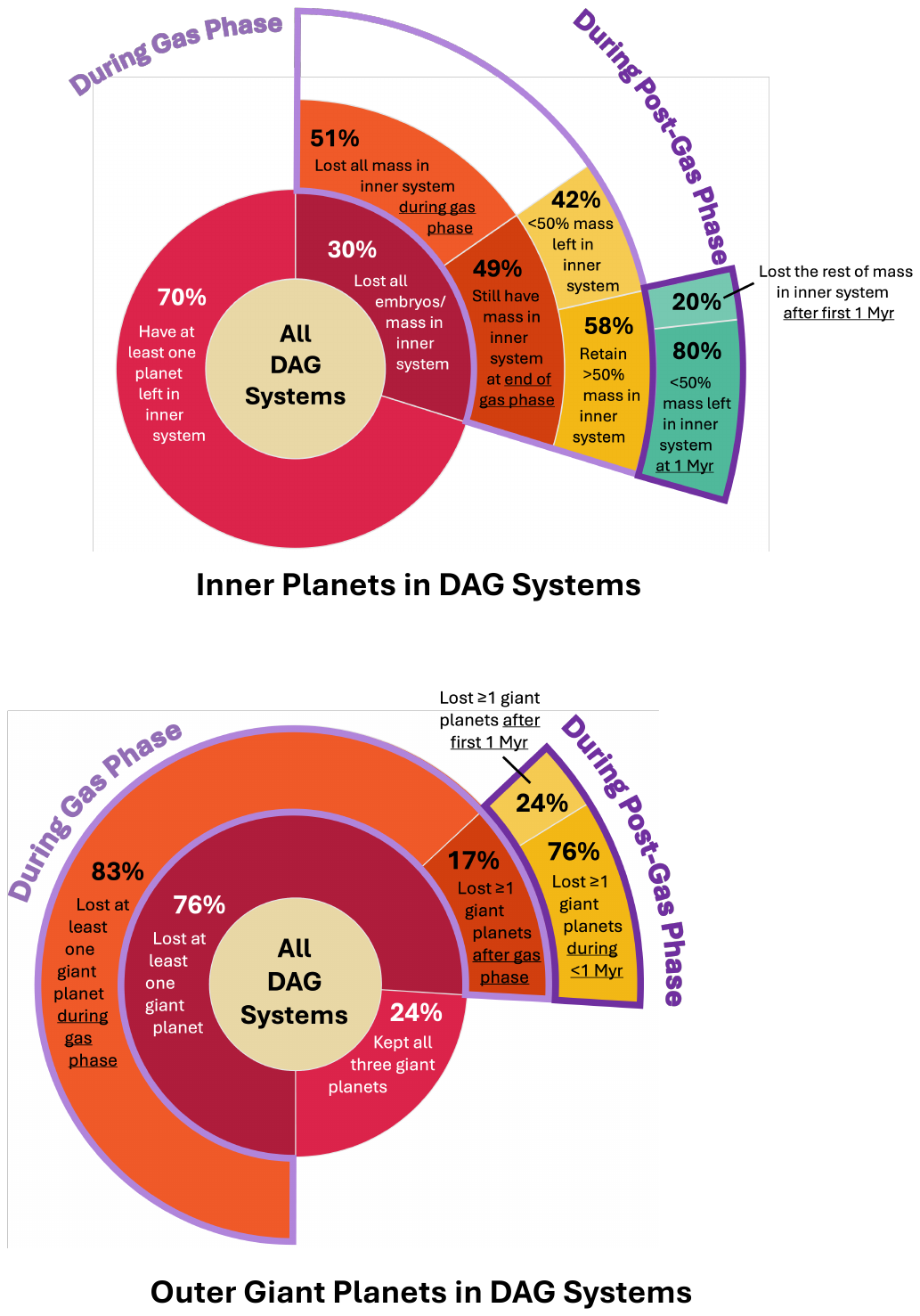}
\epsscale{3}
\caption{Sunburst diagrams describing the evolution of the inner planets (top) and the outer giant planets (bottom) in the DAG simulations. \label{fig:DAGSystems_Diagrams}}
\end{figure*}

The eccentricity excitation and scatterings caused by the giant planets reduce the number of small planets that form (Figure \ref{fig:NumSurvivingPlanets}).  In about 30\% of simulations, no embryos or planets remain interior to 1 AU.  The destruction of the inner system in these simulations typically occurs within the first 2 Myr of integration. During the gas disk stage, about half of these empty inner systems lose their embryos or planets and about 75\% lose at least half of their initial mass within 1 AU.  The majority of the remaining half of empty systems lose their inner systems in the first 1 Myr of the post-gas stage.  These empty inner systems typically emerge with one or two outer giant planets on moderately to highly elliptical orbits. Figure \ref{fig:DAGSystems_Diagrams} shows a visualization of this evolution of the forming inner planets in the DAG systems.  

The giant planets also undergo their most dramatic evolution early on. Among the 76\% of systems that lose one or more giant planets through ejections or mergers, 85\% lose at least one giant planet during the gas disk stage, and 96\% have lost at least one giant planet by 1 Myr into the post-gas stage.  Figure \ref{fig:DAGSystems_Diagrams} illustrates this evolution of the outer giant planets in the DAG systems.  The number of surviving giant planets at the end of the DAG simulations is largely determined by the initial spacing and mass of the dynamically active giant planets, with closer spacings and higher total masses resulting in fewer surviving giant planets.  This connection between the initial and final configurations of giant planets aligns with the fact that the majority of the their dynamical evolution occurs early on.

\begin{figure}[htbp]
\plotone{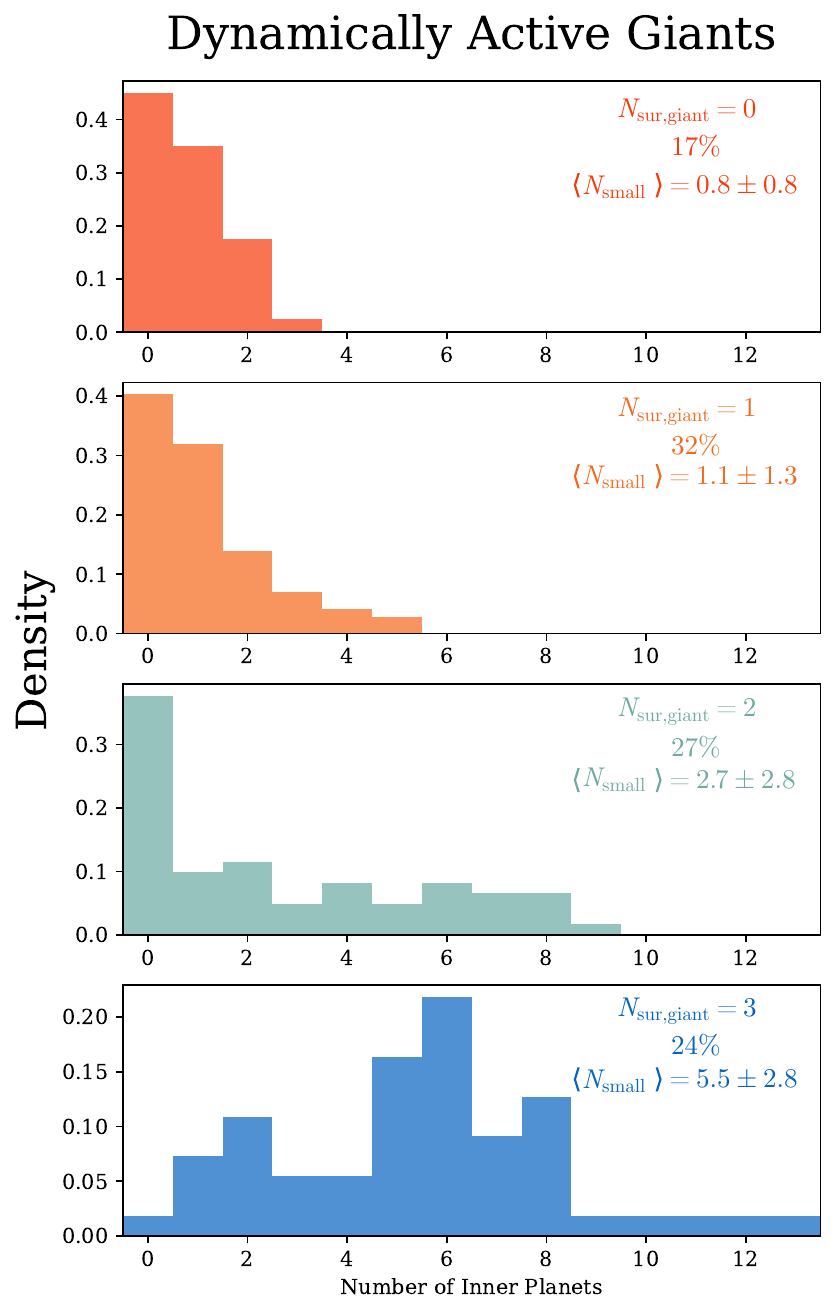} 
\caption{Histograms of the number of inner planets ($\Nsp$) formed in the DAG systems where: a)  zero giant planets survived ($\Nsg=0$); b) $\Nsg=1$; c) $\Nsg=2$ systems; and d) all three giant planets survived ($\Nsg=3$).  The $\Nsg=3$ systems form many more inner planets than those where even one giant planet is ejected. Labels indicate the fraction of systems with $\Nsg$ and the mean and standard deviation for $\Nsp$. \label{fig:unstableGiantsNumSurvivingPlanets}}
\end{figure}

In our simulations, systems that lose one or more giant planets form fewer inner planets. Figure \ref{fig:unstableGiantsNumSurvivingPlanets} shows how the number of small planets ($\Nsp$) formed in the dynamically active giant systems changes with the number of surviving giant planets ($\Nsg$).   Systems with zero or one surviving giant planet form on average one inner planet. Systems with two surviving giant planets form a mean of 2.7 inner planets. Systems where all three giant plants survive form a mean of 5.5 inner planets. 

\begin{figure}[htbp]
\plotone{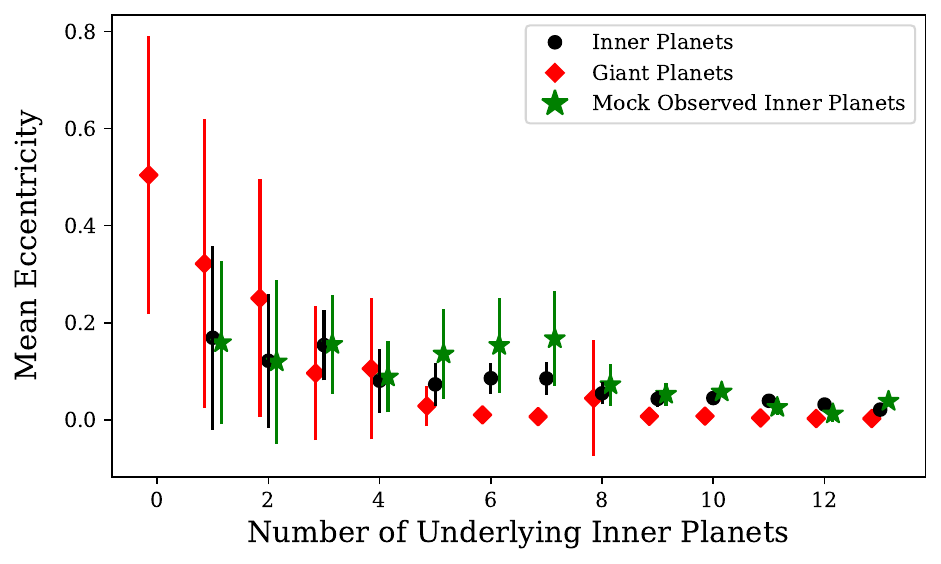}
\plotone{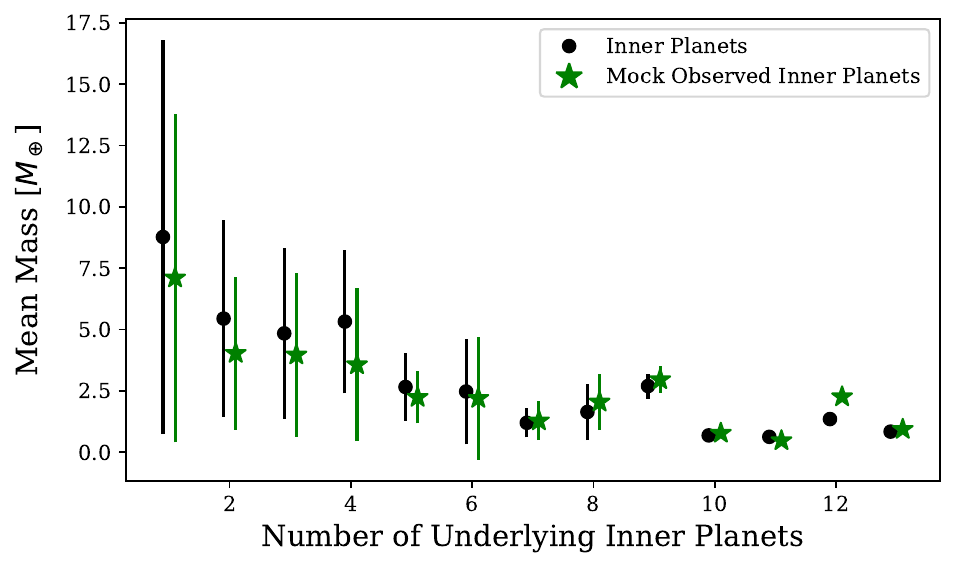}
\caption{Top: Mean intrinsic eccentricity for inner planets (black points) and giant planets (red diamonds) as a function of number of inner planets formed. The green stars indicate the mean of the mock observed sample for planets that came from an underlying system with a particular number of planets.
Bottom: Same for planet mass.
\label{fig:vsinner}}
\end{figure}

Systems that form a small number of inner planets are typically accompanied by elliptical giant planets. Figure \ref{fig:vsinner} shows the mean eccentricity for giant and small inner planets (top) and the mean mass of inner planets (bottom) as a function of the number of small inner planets formed. Systems with fewer small inner planets are the products of more mergers and tend to have higher average masses. The mean eccentricity of the surviving giant planets declines steeply with the number of inner planets for $\Nsp \le 2$. In systems that form three or more small inner planets, the surviving giant planets have low eccentricities of $\lesssim 0.1$. 

To summarize, compared to the NGP simulations, the DAG simulations produced:

\begin{itemize}
    \setlength\itemsep{0.2em}
    \item Inner planets with higher eccentricities and inclinations
    \item A population of planets within the initial disk truncation region ($\lesssim 0.05$ AU)
    \item Fewer, but more massive planets
\end{itemize}

\subsubsection{Comparison to Other Studies}

Similar to other studies using $N$-body simulations \citep[e.g.,][]{Huang2017, MustillDavies2017_MNRAS, Bitsch2023}, we find that the dynamically active giant planets reduce the intrinsic multiplicity compared to simulations without giant planets. Our resulting inner planet multiplicities are comparable to \citet{Huang2017}'s simulations that begin with three fully-formed inner super-Earth systems. In their fiducial simulations, 38\% of systems retain no inner super-Earths, 20\% retain one inner super-Earths, and 42\% retain two or more inner super-Earths. In our simulations, 30\% form no inner planets, 30\% form one, and 40\% form two or more. Of course, our systems reach higher multiplicities (up to 13), whereas \citet{Huang2017}'s simulations can retain at most three.  \revise{Our distribution of DAG inner planet multiplicities (bottom panel of Figure \ref{fig:NumSurvivingPlanets}) resembles the general shape of the unstable systems from \citet{Bitsch2023}, where the distribution peaks at having no surviving inner planets and decreases rapidly with increasing multiplicity.}

As seen in Figure \ref{fig:vsinner}, the lower the multiplicity of inner planets, the higher the mean eccentricity of the giant planets still in the system. This trend is \revise{also seen in simulations beginning with fully formed super-Earths \citep[e.g.,][]{Huang2017}, as well as simulations that include pebble and gas accretion and planetary migration \citep[e.g.,][]{Bitsch2023}.}

Systems of one or two small inner planets exhibit a wide range of eccentricities (Figure \ref{fig:vsinner}: black points) and inclinations for the inner planets, reflecting different outcomes of the competition between excitation via scattering or giant planets' secular influence vs.\ damping via mergers. Compared to \citet{Huang2017}'s simulations that begin with three fully-formed inner super-Earth systems and that have no gas damping and fewer mergers, our inner planets have somewhat lower eccentricities and inclinations. The surviving single super-Earths in their fiducial simulations have a mean eccentricity of 0.4 and mean inclination of $30^\circ$, whereas our single inner planets have a mean eccentricity of 0.18 and mean inclination of 17$^\circ$. 

\begin{figure}[htbp]
\plotone{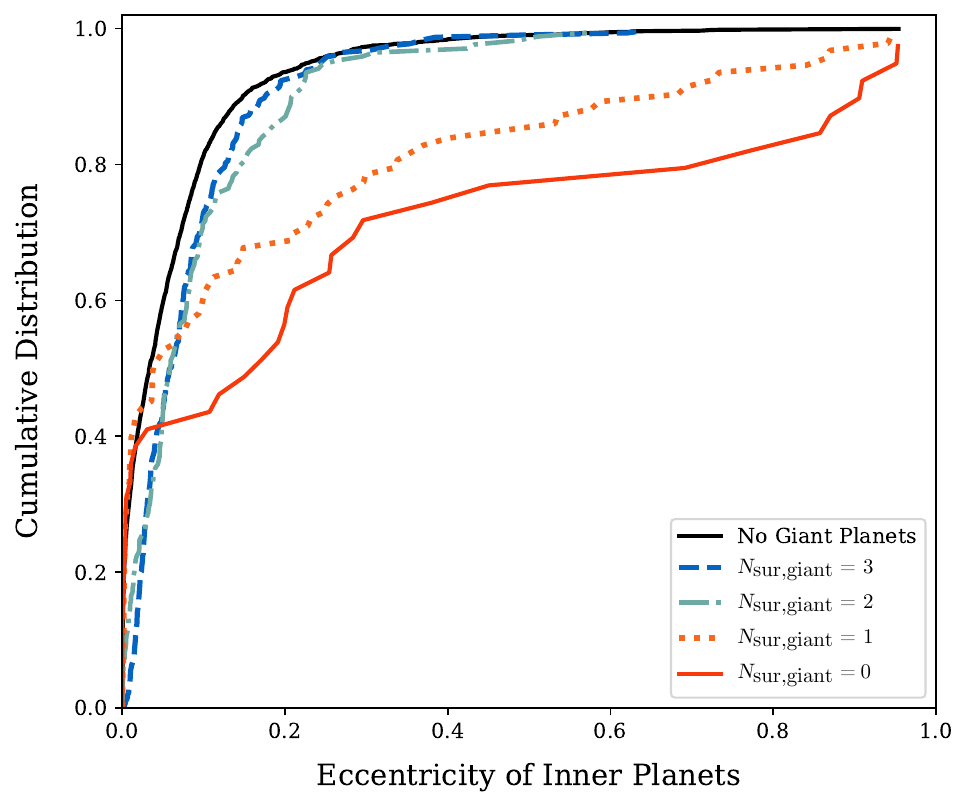}
\epsscale{1.3}
\caption{Cumulative distribution of inner planet eccentricities for DAG systems, broken down by number of surviving giant planets, \Nsg.  The eccentricity distribution for the NGP simulations (solid black line) is plotted for comparison.  \label{fig:CDF_Eccs_Nsg}}
\end{figure}

We also can break down the eccentricity distributions of the inner planets by number of surviving giant planets.  Figure \ref{fig:CDF_Eccs_Nsg} shows how the DAG inner planets' eccentricity distributions compare depending on number of surviving giant planets.  Similar to \citet{MustillDavies2017_MNRAS}, we find that the excitement of inner planet eccentricities increases as a system loses more of its outer giant planets.  Additionally, we note that even when all of the dynamically active giant planets survive (\Nsg $=3$), the inner planets' eccentricities are still elevated as compared to the NGP eccentricities (Figure \ref{fig:CDF_Eccs_Nsg}: solid black line).

\subsection{Effects on observed multiplicity and spacing \label{subsec:effectObsMult_Spacing}} 

\begin{figure*}[htbp]
\epsscale{1.2}
\plotone{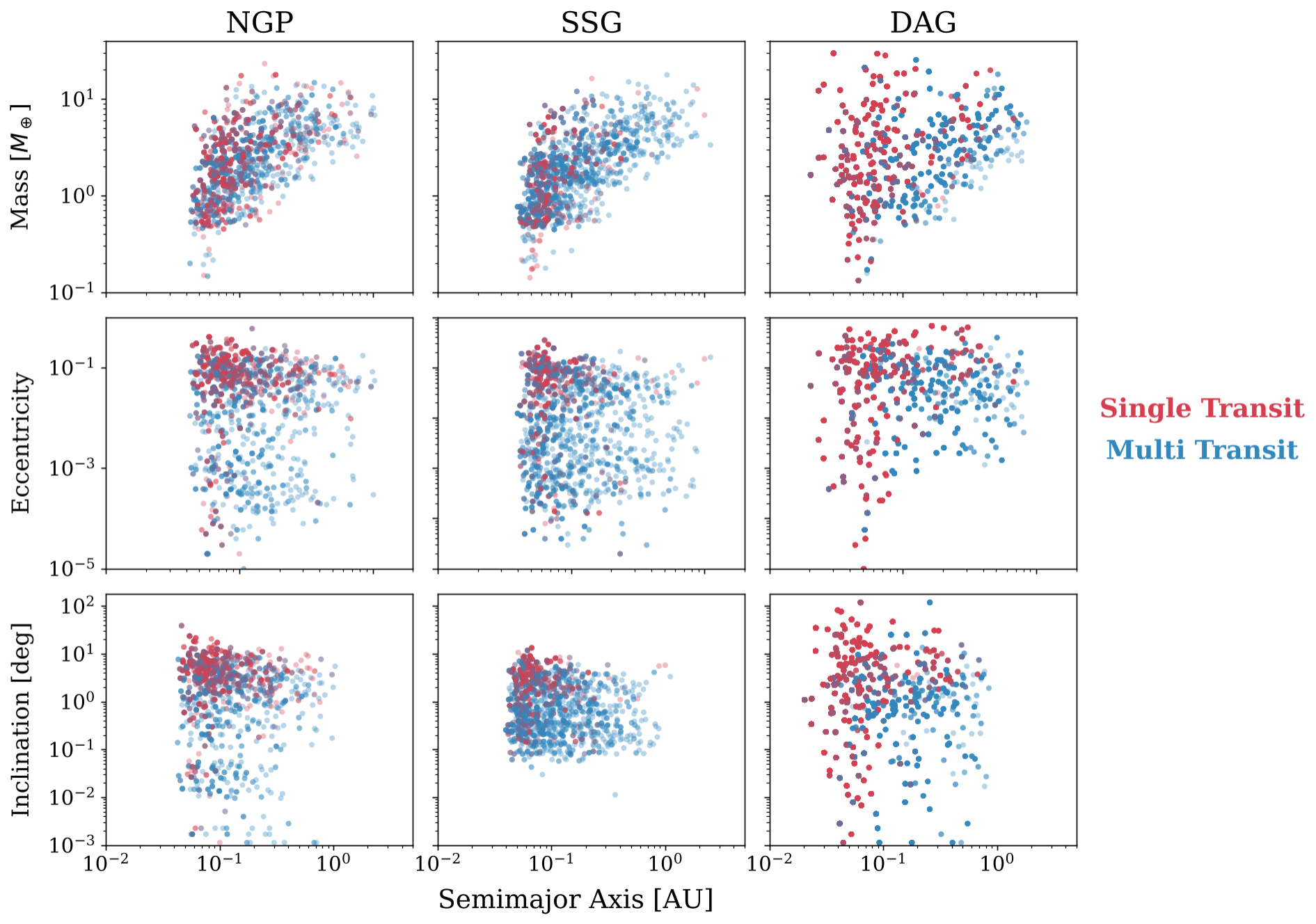}
\caption{Mass (top row), eccentricity (middle row), and inclination (bottom row) vs.\ semimajor axis of mock observed planets that are detected are as singles (red) or multis (blue) from the NGP (left column), SSG (middle column), and DAG (right column) simulations. \revise{The inclinations were measured with respect to the reference plane in \texttt{REBOUND}, which corresponds to the $x$-$y$ plane.} \label{fig:observed}}
\end{figure*}

As described in Section \ref{subsec:forwardModeling}, we forward model transit detections to compare to the \textit{Kepler} candidates. Figure \ref{fig:observed} shows the masses, eccentricities, and inclinations of our mock detected planets. Comparing Figure \ref{fig:observed} to the bottom row of Figure \ref{fig:time}, we tend to detect close-in planets with a large range of masses and at larger semimajor axes, we detect only higher mass planets.  This general trend of the planets detected at larger semimajor axes having higher masses is expected due to the incorporation of selection effects in our forward modeling transit detection pipeline (described in detail in Section \ref{subsec:forwardModeling}).

\begin{figure*}[htbp]
\plotone{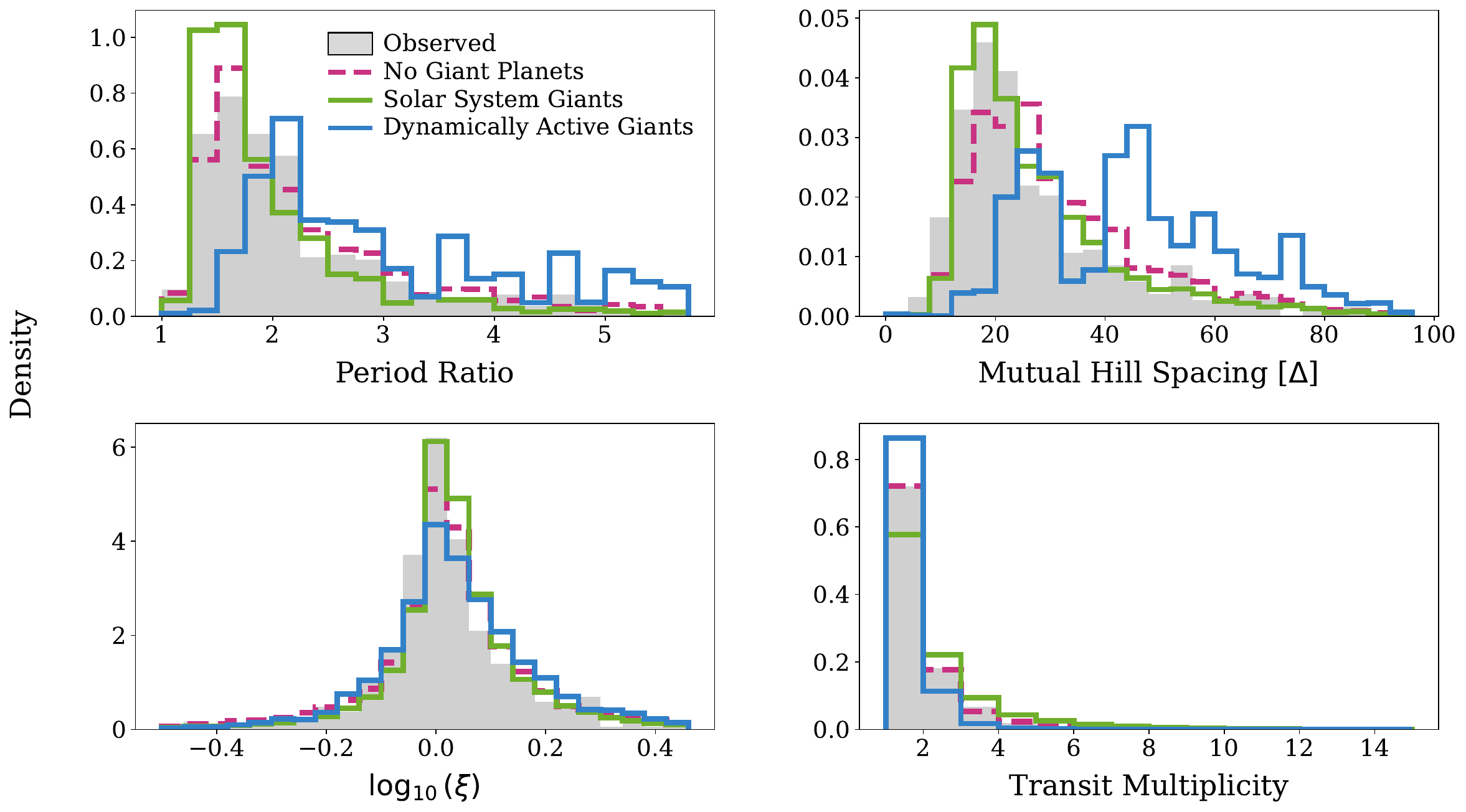} 
\caption{Mock observed period ratio of adjacent planets (top left), Hill spacing between adjacent planets (top right), ratio of transit duration between adjacent planets (bottom left), and the transit multiplicity (bottom right).  The solid gray histogram corresponds to the observed \textit{Kepler} distributions; the dashed pink line corresponds to the simulations without outer giant planets inserted (data from \citealt{MacDonald2020}); the solid green and blue lines correspond to the simulations with the Solar System giant planets and the dynamically active giant planets, respectively.
\label{fig:observableDistributions}}
\end{figure*}

Figure \ref{fig:observableDistributions} compares four forward-modeled observables. Compared to the NGP simulations, the increased scatterings and mergers in the DAG simulations lead to wider spacings in period ratio and mutual Hill spacing, more single transiting planets, and a wider and more symmetric distribution of transit duration ratios ($\log_{10} \xi$). The wider $\log_{10} \xi$ distribution reflects the larger eccentricities \revise{(Figure \ref{fig:observed})} and mutual inclinations.

There are several contributors to the excess of single transiting planets in the DAG simulations. The most significant contributor is the large number of systems containing only one planet interior to 1 AU (Figure \ref{fig:NumSurvivingPlanets}), which make up 31\% of the mock observed single transiting planets. These single inner planet systems also have higher masses (Figure \ref{fig:vsinner}) and---under our assumed mass-radius relationship---larger radii, making them easily detectable and overrepresented in the mock observed sample. Another contributor is the higher mutual inclinations (Figure \ref{fig:time}, bottom left and Fig. \ref{fig:observed}, bottom right). Among transiting planets with high mutual inclinations $(>3^\circ)$ with respect to their underlying adjacent neighbors, 72\% are observed as singles, whereas only 52\% of planets with lower mutual inclinations are observed as singles. The final contributor is the population of planets with $a<0.05$ AU that does not exist within the no giant planet simulations; 81\% of these mock observed planets are singles, compared to 69\% of mocked observed planets with $a>0.05$ AU. 

Compared to \citet{Huang2017}'s simulations, we generate more systems with low transit multiplicity. Their ratio of two transiting to single transit systems is 0.2 for their fiducial simulations, while ours is 0.1. The simulated planets in \citet{Huang2017} have a higher fraction with only one intrinsic planet, high mutual inclinations, and include planets with $a<0.05$ AU, so the factors above are likely not the explanation. Instead, we hypothesize that we produce more singles due to incorporating detection efficiency, resulting in systems where only one planet is detected even though more geometrically transit, especially given that our simulations extend to lower planet masses and assumed radii and therefore lower signal-to-noise ratios.

\subsection{Effects on observed eccentricities}

\begin{figure}[htbp]
\plotone{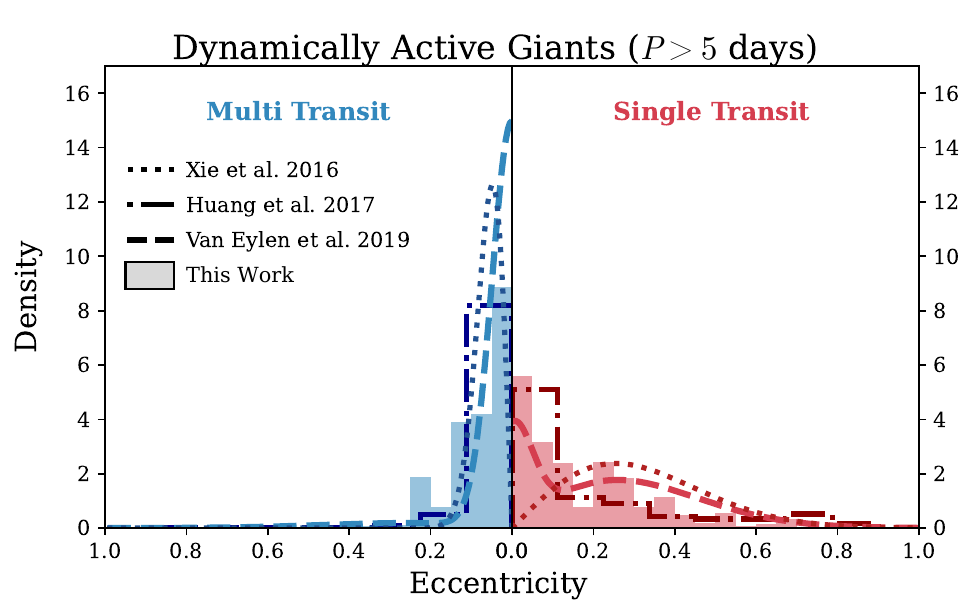} 
\plotone{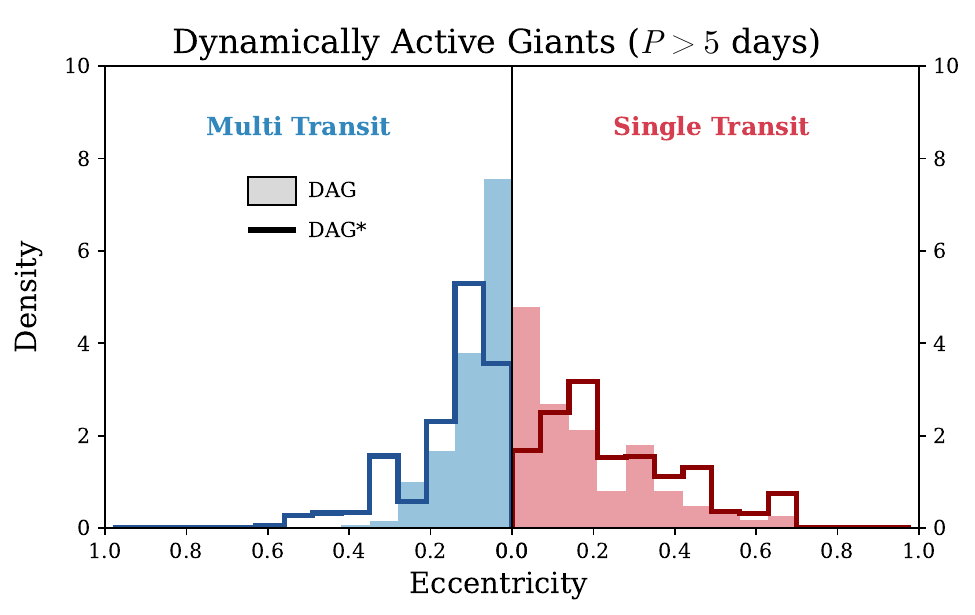} 
\plotone{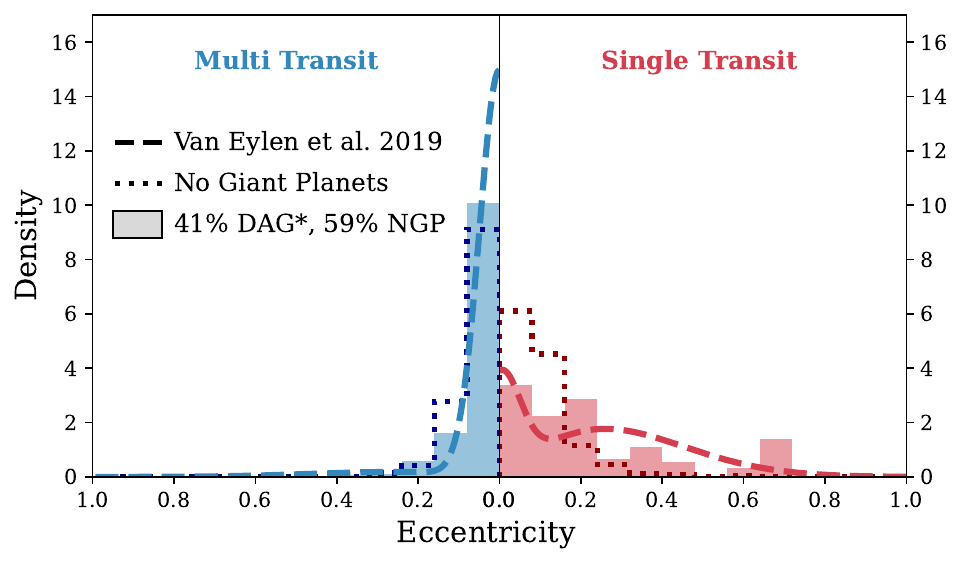} 
\caption{Eccentricity distribution of multi (left, in blue) and singly (right, in red) transiting systems. Top: The DAG mock observed planets are shown as solid histograms with the inferred observed distribution from \citet{XieDong2016_PNAS} overplotted as a dotted line, inferred observed distribution from \citet{VanEylen2019} plotted as a dashed line, and simulated distribution from \citet{Huang2017} overplotted as a dot dashed line. Our simulated distribution and the \citet{VanEylen2019} exclude planets with periods less than 5 days. Middle: Mock observed planets from simulations with dynamically active giant planets with (DAG, shaded) vs.\ without (DAG*, solid line) a gas damping stage. Bottom: The shaded histogram shows our mixture where $41\%$ of mock observed systems are accompanied by outer giants from DAG* simulations (DAG simulations without gas damping) and $59\%$ are from simulations without giant planets. \label{fig:edist}}
\end{figure}

In Figure \ref{fig:edist}, we plot the distributions of eccentricities for the single and multi transiting planets with orbital periods beyond 5 days. Following \cite{VanEylen2019}, we limit our sample to planets beyond 5 days to exclude planets likely to be tidally circularized. For single transiting planets, our DAG distribution has a mean $\overline{e}_\text{single} = 0.17$, which is larger than the NGP distribution ($\overline{e}_\text{single} = 0.10$), but smaller than observed $\overline{e}_\text{single} = 0.32 \pm 0.02$ from \citet{XieDong2016_PNAS} or $\overline{e}_\text{single} = 0.25^{+0.07}_{-0.06}$ from \citet{VanEylen2019}. Our DAG standard deviation $\sigma_{e,\text{single}} = 0.16$ is consistent with \citet{XieDong2016_PNAS}'s $\sigma_{e,\text{single}} = 0.17 \pm 0.10$ and \citet{VanEylen2019}'s $\sigma_{e,\text{single}} = 0.19^{+0.07}_{-0.06}$. The DAG single transiting planet mean is consistent with \citet{Huang2017}'s simulations that begin with fully-formed inner super-Earth systems and dynamically active outer giants, which have $\overline{e}_\text{single} = 0.19 \pm 0.03$.  However, \citet{Huang2017} better produce the tail of very large eccentricities (Figure \ref{fig:edist}, top row). Our deficit of high eccentricities is likely due to the collisional and gas damping of eccentricities as the embryos grow, allowing them to reach more circular configurations that persist through the post gas stage. 

To explore the role of the gas damping stage, we analyze a suite of otherwise identical DAG simulations that do not include gas damping, which we call DAG*. \reviseTwo{By simulating systems without the 1~Myr integration within a depleted gas disk, we assume that these systems have their inner system embryos reach isolation mass after the gas disk has fully dissipated.} The resulting planets exhibit higher eccentricities for singles ($\overline{e}_\text{single} = 0.26$ and $\sigma_{e,\text{single}} = 0.17$) that more closely match observed values. They also better match the tail of large eccentricities (Figure \ref{fig:edist}, middle row).

For multi-transiting systems, the DAG* simulations produce eccentricities that are too high. Our means are $\overline{e}_\text{multi} = 0.06$ from the mock observed NGP simulations and $\overline{e}_\text{multi} = 0.16$ from the mock observed DAG* simulations, compared to the observed $\overline{e}_\text{mult} = 0.04^{+0.02}_{-0.04}$ from \citet{XieDong2016_PNAS} and $\overline{e}_\text{mult} = 0.06\pm0.02$ from \citet{VanEylen2019}. \citet{Huang2017}'s simulations produce $\overline{e}_\text{mult} = 0.04\pm0.02$.  Our standard deviations are $\sigma_{e,\text{multi}} = 0.06$ for the mock observed NGP simulations and $\sigma_{e,\text{multi}} = 0.12$ for the mock observed DAG* simulations, compared to the observed $\sigma_{e,\text{multi}}  = 0.08^{+0.03}_{-0.05}$ from \citet{VanEylen2019}. Even our NGP simulations' $\sigma_{e,\text{multi}}$ is somewhat larger than the observed $\sigma_{e,\text{multi}}  = 0.021^{+0.010}_{-0.021}$ from \citet{XieDong2016_PNAS}.

\section{Solar System Giant Planet Simulations}
\label{sec:ssg}
 Compared to both the DAG and NGP simulations, the Solar System Giants, or SSG, simulations tend to produce systems that are lower in mass, higher in intrinsic multiplicity, and dynamically colder. In the middle panel of Figure \ref{fig:underlying}, we can see the underlying systems are more compact than systems from the NGP simulations (left panel). The histogram in the middle panel in Figure \ref{fig:NumSurvivingPlanets} shows that the simulations form more inner planets in each system. Figure 
\ref{fig:observableDistributions} shows that the forward-modeled mock transiting planets have smaller period ratios and transit multiplicity, somewhat smaller mutual Hill spacings, and more duration ratios closer to 1 than the NGP simulations. Table \ref{tab:properties} summarizes the properties of these systems, including their lower eccentricities and mutual inclinations. Since the Solar System giant planets are initially in nearly coplanar, widely separated, circular orbits, we suspect that their secular effect on inner forming planets---particularly the outermost inner planets---keeps the inner systems well-aligned. These findings motivate continuing observational work to characterize the orbital properties of outer companions to inner planetary systems and correlate outer and inner planet properties.

The trends in underlying planets have some similarities to those found by \citet{ChildsQuintana2019_MNRAS} in their simulations, which focused on Earth analogs formed in lower mass (total mass 4.85 $M_\oplus$), wider (embryos spaced from 0.35 to 4 AU) disks. \citet{ChildsQuintana2019_MNRAS} compare simulations with Jupiter and Saturn analogs to those with lower mass planets. Their simulations with Jupiter and Saturn analogs also result in lower eccentricities and inclinations for the inner planets. However, in \citet{ChildsQuintana2019_MNRAS}'s simulations, ejection of embryos initially located beyond 1 AU truncated the semimajor axis distribution. This truncation is not a significant effect for our simulations, in which all of the embryos were initially located within 1 AU. 

\section{Comparison of Outcomes from Different Configurations of Giant Planets}
\label{sec:comparison}

We can also directly compare the outcomes from our simulations with two different outer giant planet configurations, a set of dynamically active giant planets (DAG) and the Solar System giant planets (SSG), as well as the outcomes from simulations \revise{from a previous work} without any outer giant planets \citep[NGP:][]{MacDonald2020}.  Table \ref{tab:properties} reports the average values of different properties of the underlying planets and the mock observed planets for each set of simulations. First, we summarize the notable differences between the DAG, SSG, and NGP cases for four properties: semimajor axis, planet mass, mutual inclination, and eccentricity.

\textbf{Semimajor axis:} The DAG simulations produce planets that are closer-in than both the SSG and NGP simulations.  This trend is seen in both the underlying distribution and the mock observed distribution.  There are fewer transiting planets beyond $a \gtrsim 0.1$ AU in the DAG systems than in the SSG or NGP systems.

\textbf{Mass:} In both the underlying and observed mass distributions, the DAG simulations produce more high mass planets, while the SSG simulations produce more low mass planets.  In the observed mass distribution, there is a larger offset between the averages of the DAG and SSG simulations.

\textbf{Mutual inclination:} The SSG systems have an excess of low mutually inclined planets and very few intermediate and high $i_\text{mut}$ compared to the NGP and DAG systems.  This excess of low mutual inclinations is likely responsible for the larger number of SSG systems that have multiple observed to be transiting.

\textbf{Eccentricity:} For both singly and multi transiting planets, the SSG systems produce planets that are observed to have lower eccentricities, while the DAG systems are observed to have higher eccentricities.

We also compare the differences in four observable properties' distributions, described in Section \ref{subsec:forwardModeling}.  The distributions of these observables are shown in Figure \ref{fig:observableDistributions}. The observed SSG systems have the smallest period ratios; whereas, the observed DAG systems have an excess of larger period ratios.  The DAG systems also produced a significant excess of mutual Hill spacings at values $\Delta \gtrsim 40$. The observed DAG systems show a slight widening of their $\log{\xi}$ distribution, indicating the presence of higher eccentricities than the SSG or NGP systems. The transit multiplicity distribution shows that the DAG systems produce an excess of singly transiting planets and the SSG systems produce more multi-transiting systems.  We previously discussed the reasons why the DAG systems generated significantly more singly transiting planets in Section \ref{subsec:effectObsMult_Spacing}.

\section{Mixing systems with and without giant planets}
\label{sec:mix}

Of course, not all super-Earth systems host giant planets, nor is it likely that all singles come from systems with giant planets and all multis from systems without them. Estimates of the fraction of observed inner super-Earths without giant planet companions vary from 10\% (occurrence rate of giant planets independent of super-Earths, e.g., \citealt{2008PASP..120..531C}) to $\sim$40\% \citep[e.g.,][]{RosenthalKnutson2022_ApJS}. Furthermore, our simulations with dynamically active giant planets are likely not representative of all giant planet systems, so the true fraction of systems that share similar evolutionary histories is likely lower. 

We choose to use our DAG simulations without a gas damping stage, which we call DAG*, instead of our original DAG simulations when mixing together populations with and without giant planets. Our motivation for using the DAG* simulations is that these simulations produce more high eccentricity singly transiting planets (Figure \ref{fig:edist} middle plot, right panel, in red) which are seen observationally.  A comparison of the observables from the DAG and DAG* simulations is in Figure \ref{fig:obsCompareDAGs} in Appendix \ref{app:mix}.  Overall, the mock observed planets produced by DAG* simulations are dynamically hotter than in the DAG simulations, which include an initial 1 Myr gas phase.

We combine our DAG* and NGP simulations to achieve a mixture where two different percentages, 10\% and 41\%, of mock observed systems are accompanied by outer giants.  In order to reflect the \textit{observational} occurrence rate of outer giant planets, we categorize the DAG* systems that lost all of their giant planets as \textit{observed} NGP systems before we mix the two populations together.

As discussed in Section \ref{subsec:forwardModeling}, we have the flexibility to weight the simulations to reflect the unknown underlying variation in the amount of solids, parameterized by $\Sigma_{z,1}$ (Equation \ref{eq:solidSurfaceDensity}), from disk to disk. Because the solids in the inner region may be influenced by the outer giant planets, we can weight the simulations with the dynamically active giant planets independently of the NGP simulations.  We choose appropriate weights for each component of the mixture by \reviseTwo{manually altering the weighting factors and performing a ``by-eye'' optimization} such that the resulting observable distributions (i.e., those shown in Figure \ref{fig:observableDistributions}) and eccentricity distributions (i.e., those shown in Figure \ref{fig:edist}) qualitatively match their respective observed distributions \reviseTwo{as much as possible}. 

\begin{figure}[htbp]
\epsscale{1.2}
\plotone{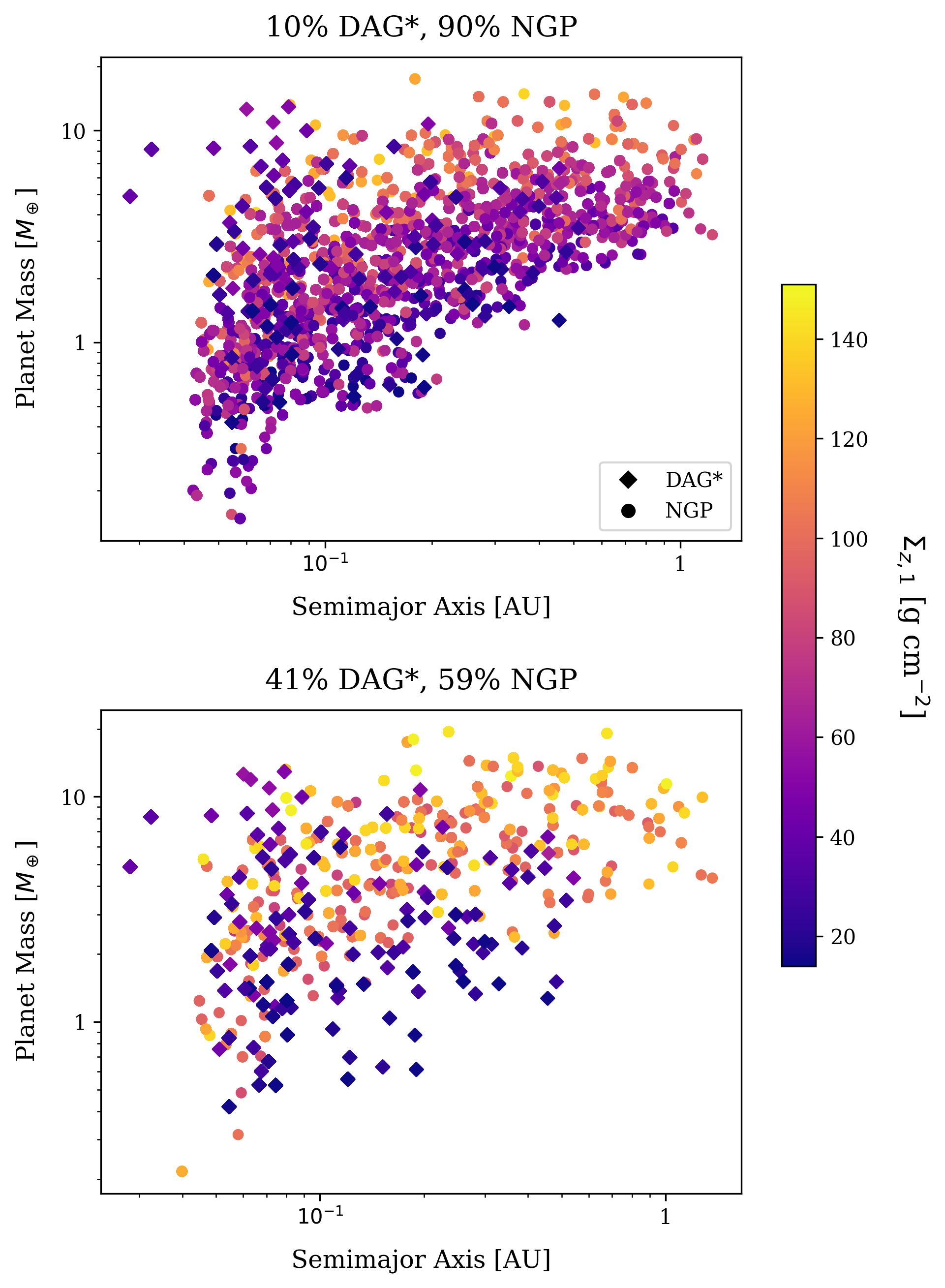} 
\caption{\revise{Plots of the mock observed inner planet masses as a function of semimajor axis for different mixtures of the DAG* simulations (DAG simulations without a gas phase) and NGP simulations.  The points are colored by the solid surface density normalization factor, representing the different formation conditions that were necessary to achieve a good qualitative match to observations.  The diamond points represent the planets that would be ``observed'' in systems with dynamically active giants; the circular points represent the planets that would be ``observed'' in systems without any giant planets.  In other words, if a system originally contained dynamically active giants but later lost these giant planets, these few systems would be counted within the NGP case.} \label{fig:mixSSDs}}
\end{figure}

For our 10\% mixture, using Equation \ref{eq:weightingFactor}, we apply the weights $\Sigma^-_{z,1} = 0 $ and $ \sigma_{\Sigma_{z,1}} = 15$ to the DAG* simulations. We apply the weights $\Sigma^-_{z,1} = 65 $ and $ \sigma_{\Sigma_{z,1}} = 20$ to the NGP simulations.  \revise{A visualization of these different solid surface density normalization factor distributions can be seen in the top panel of Figure \ref{fig:mixSSDs}.} The resulting distributions of observables (Appendix \ref{app:mix}) are similar to the nominal NGP case\footnote{\revise{We experimented with a large range of different weights for both components of this mixture; however, all combinations of chosen weights produced negligible changes to the observable distributions, as compared to the NGP case.  The adopted weights for this mixture, as listed above, altered the observable distributions \textit{the most} of those tested.}}. Since the inclusion of systems with outer giants does not significantly alter the resulting distributions of observables, this implies at least 10\% of transiting super-Earths could be accompanied by outer giants. However, this modest fraction does not improve our ability to match the observed high eccentricity tail. 

Next we pool our DAG* and NGP simulations to achieve a mixture where 41\% of mock observed systems are accompanied by outer giants.  Using Equation \ref{eq:weightingFactor}, we apply the weights $\Sigma^-_{z,1} = 0 $ and $ \sigma_{\Sigma_{z,1}} = 15$ to the DAG* simulations. We apply the weights $\Sigma^-_{z,1} = 130 $ and $ \sigma_{\Sigma_{z,1}} = 15$ to the NGP simulations. \revise{A visualization of these different solid surface density normalization factor distributions can be seen in bottom panel of Figure \ref{fig:mixSSDs}.} We now produce a good qualitative match to the period ratio, mutual Hill spacing, duration ratio, and transit multiplicity (Appendix \ref{app:mix}), while also producing more observable elliptical inner planets (Figure \ref{fig:edist}, bottom row). For this mixture,  we find $\overline{e}_\text{single} = 0.22$, $\sigma_{e,\text{single}} = 0.20$, $\overline{e}_\text{multi} = 0.04$, and $\sigma_{e,\text{multi}} = 0.07$. This 41\% mixture accounts for the observed higher eccentricity singles (Figure \ref{fig:underlying}, bottom row, right, in red) while matching the observed dynamically cold population (Figure \ref{fig:underlying}, bottom row, left, in blue, and Appendix \ref{app:mix}).

\section{Summary and Discussion} \label{sec:prelimConclusionsandFutureWork}
With increasingly longer and more precise surveys and instrumentation, studies have begun to quantify the conditional occurrence rates of different classes of planets (e.g., super-Earths and outer giant planets).  However, the sample of small inner planet systems with known outer giant companions is not yet large enough to quantify how these conditional occurrence rates depend on other planetary properties, such as multiplicity, eccentricity, or spacing. Additionally, there are still several aspects of observed properties of small inner planets (e.g., excess of singly transiting planets, eccentricity dichotomy) that have not been comprehensively explained via proposed formation mechanisms.  We aimed to explore possible connections between the properties of inner and outer systems and how different initial giant planet configurations could affect how formation proceeds and changes both underlying and observed planetary properties.   

We performed $N$-body simulations of late-stage in situ planet formation in the presence of outer giant planets.  We investigated the effects on the final system configurations from two sets of giant planets: the four Solar System giant planets and three dynamically active giant planets.  The presence of the giant planets affected the underlying architecture of the systems and produced measurable differences in the distributions of observable properties.  

Compared to simulations without any outer giants (NGP) and simulations with the Solar System giants (SSG), the simulations with dynamically active giant planets (DAG) tended to produce systems that were dynamically hotter: (1) larger mutual Hill spacings; (2) higher eccentricities; (3) an excess of single transits; (4) larger period ratios; (5) planets closer to the central star; and (6) fewer, but more massive, planets.  

Compared to the DAG and NGP simulations, the SSG simulations tended to produce systems that were dynamically colder: (1) smaller mutual Hill spacings; (2) higher transit multiplicities; (3) lower eccentricities; (4) smaller period ratios; (5) smaller mutual inclinations; (6) more surviving planets; and (7) less massive planets.

These results broadly agree with earlier studies that explored different formation and evolution scenarios, including excitation of eccentricities and reduction in transit multiplicity by outer giant planets introduced into mature planetary systems \citep{Huang2017} and into young systems undergoing formation and migration \citep{Bitsch2023}, and reductions in eccentricity and mutual inclinations for Earth-like planet forming in the presence of more distant outer giants \citep{ChildsQuintana2019_MNRAS}.

Including a contribution from systems that form with dynamically active giant planets helps resolve a prior discrepancy between in situ formation models for inner super Earths and the observed Kepler population. \citet{MacDonald2020} found that simulations of in situ formation via giant impacts can account for observed properties like period ratio and transit multiplicity, but this model did not account for the observed high eccentricities of some single transiting planets \citep{VanEylen2019}. We found that a mix of simulations formed with and without giant planets, combined so that 41\% of mock observed inner systems reside in systems containing giant planets, can simultaneously account for the observed eccentricities of single and multi-transiting system along with period ratio, duration ratio, Hill spacing, and transit multiplicity. 

However, matching the observed population required different formation conditions prior to the giant impact stage for systems with vs.\ without giant planets. Our non-giant planet component primarily formed from disks with high solid surface densities and underwent many of their giant impacts in a depleted gas disk. Our giant planet component (i.e., super-Earths forming in systems with dynamically active giant planets) underwent giant impacts without a depleted gas disk (i.e., where giant impacts commence in a gas-free environment) and in disks with lower solid surface density, conditions that are more similar to those assumed for the solar system terrestrial planets. 

\reviseTwo{Given the observed correlations between higher stellar metallicities and the occurrence rate of giant planets \citep[e.g.,][]{SantosIsraelian2004_A&A, FischerValenti2005_ApJ}, it may be reasonable to assume that systems containing giant planets should on average initially have larger reservoirs of solids in their protoplanetary disks.  We did not directly assume any underlying correlations and, instead, explored the full range of solid surface densities.  Since the DAG* systems with lower solid surface densities produced the dynamically hot planets necessary to match observations, this} suggests that giant planets may have inhibited---but not entirely prevented---delivery of solid material to the inner disk. Another possibility is that the giant planets inhibited the incorporation of solid material into embryos, for example, by influencing parameters like pebble size (e.g., \citealt{RMC2020}) that may affect pebble isolation mass. The formation without a depleted gas disk would suggest that giant planets influence the inner gas disk, perhaps clearing a cavity more rapidly than photoevaporation alone.

%\reviseTwo{add paragraph here about metallicity}

Although our mixed population scenario\revise{s are} broadly consistent with the observed occurrence of giants (e.g., \citealt{RosenthalKnutson2022_ApJS}), \revise{we recognize that the DAG configuration used in matching the observational distributions is  unlikely to be the primary component of the formation histories of systems with outer giant planets.  However, we have demonstrated that a mixture with 10\% of systems coming from the DAG* simulations, which resulted in extremely dynamically hot planetary properties, did little to change any of the observable distributions as compared to simulations without any giant planets (see Figures \ref{fig:edist10} and \ref{fig:obs10}).} 

\revise{Going forward,} we recommend further observational and modeling work to assess \revise{the occurrence rates of outer giant planets} in more detail. A wider range of configurations for giant planets should be explored with simulations to better delineate what initial configurations dynamically heat vs.\ dynamically cool the inner systems. A larger sample size for occurrence rates could allow observers to better probe inner super-Earth occurrence rate as a function of outer giant planet properties like eccentricity, multiplicity, and spacing.  This larger sample will allow a better comparison to simulation results and could further break down super-Earth occurrence rates by inner planet properties like size and multiplicity. Ultimately we need to inventory a subset of planetary systems with as much completeness as possible to better study the connections among these different components.  \revise{Uniting a comprehensive sample of planetary systems with modeling how different configurations of outer giant planets affect the formation and dynamical evolution of terrestrial planets will be a crucial step in understanding the broad landscape of habitability.}

%\begin{acknowledgments}
    The Center for Exoplanets and Habitable Worlds is supported by the Pennsylvania State University, the Eberly College of Science, and the Pennsylvania Space Grant Consortium.  PHS and RID were supported in part by NASA Exoplanet Research Program grant No. 80NSSC24K0150. SJM was supported in part by a Faculty Research Grant at Missouri State University and the NASA-Missouri Space Grant No. 80NSSC20M0100. Computations for this research were performed on the Pennsylvania State University’s Institute for Computational and Data Sciences’ Roar supercomputer
  %  \end{acknowledgments}

\software{rebound \citep{rebound:paper2012, rebound:paper2015a, rebound:paper2015b, rebound:paper2019}, reboundx \citep{reboundx:paper}, arviz \citep{arviz:paper, arviz:zenodo}, astropy \citep{astropy:paper2013, astropy:paper2018, astropy:paper2022, astropy:zenodo}, forecaster \citep{forecaster:paper}, h5py \citep{h5py:paper}, matplotlib \citep{matplotlib:paper, matplotlib:zenodo}, numpy \citep{numpy:paper}, pandas \citep{pandas:paper, pandas:zenodo}, pyreadstat \citep{pyreadstat:zenodo}, scipy  \citep{scipy:paper, scipy:zenodo}}

\appendix

\section{Mixtures of Dynamically Active Giant Planets and No Giant Planets Simulations}
\label{app:mix}

\begin{figure}[htbp]
\plotone{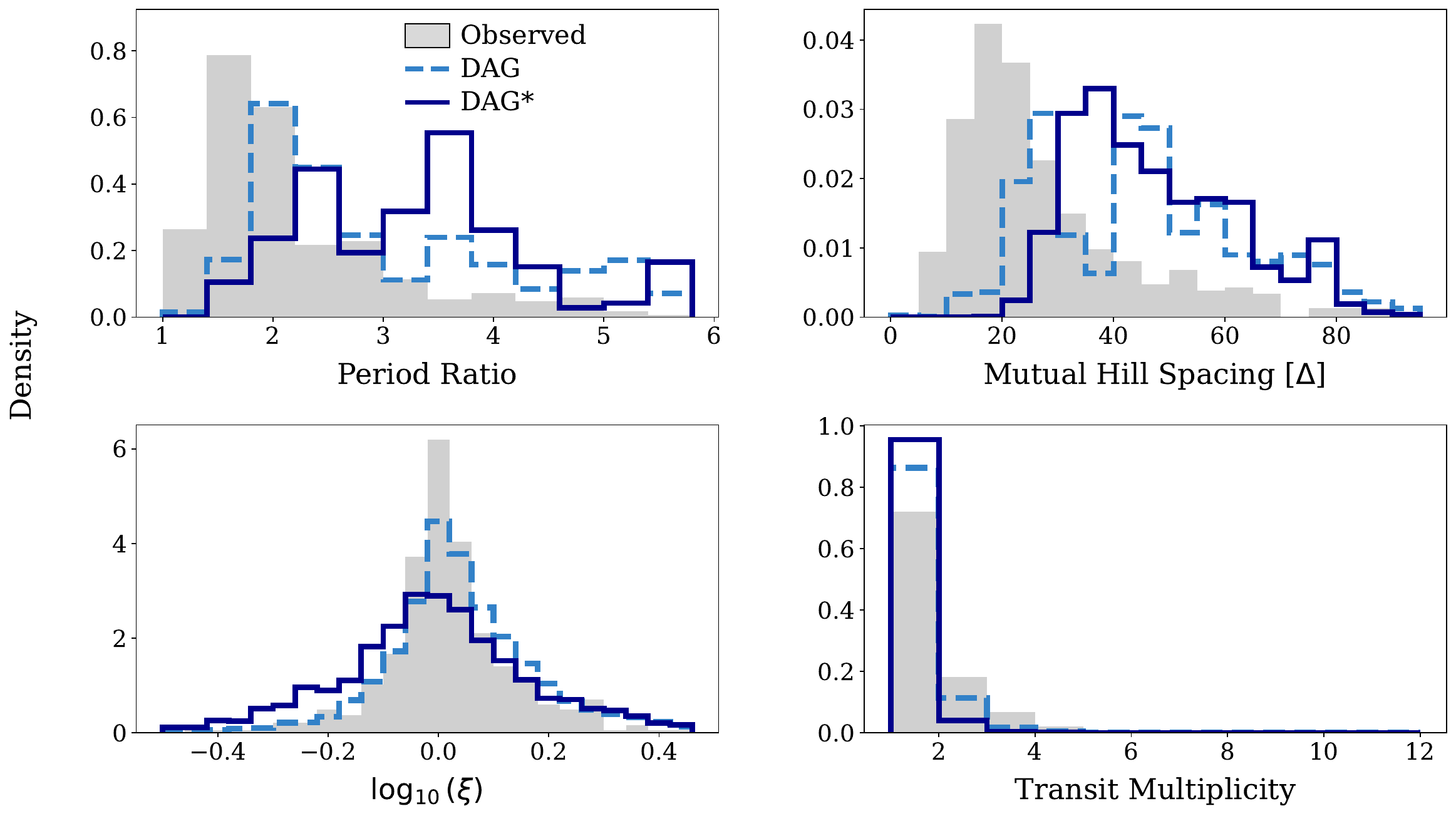} 
\caption{Comparison of the observable distributions for the DAG and DAG* simulations. Mock observed period ratio of adjacent planets (top left), Hill spacing between adjacent planets (top right), ratio of transit duration between adjacent planets (bottom left), and the transit multiplicity (bottom right).  The solid gray histogram corresponds to the observed \textit{Kepler} distributions; the dashed light blue line corresponds to the original DAG simulations; and the solid dark blue line corresponds to the DAG* simulations (DAG simulations without a gas phase).\label{fig:obsCompareDAGs}}
\end{figure}

\begin{figure}[htbp]
\plotone{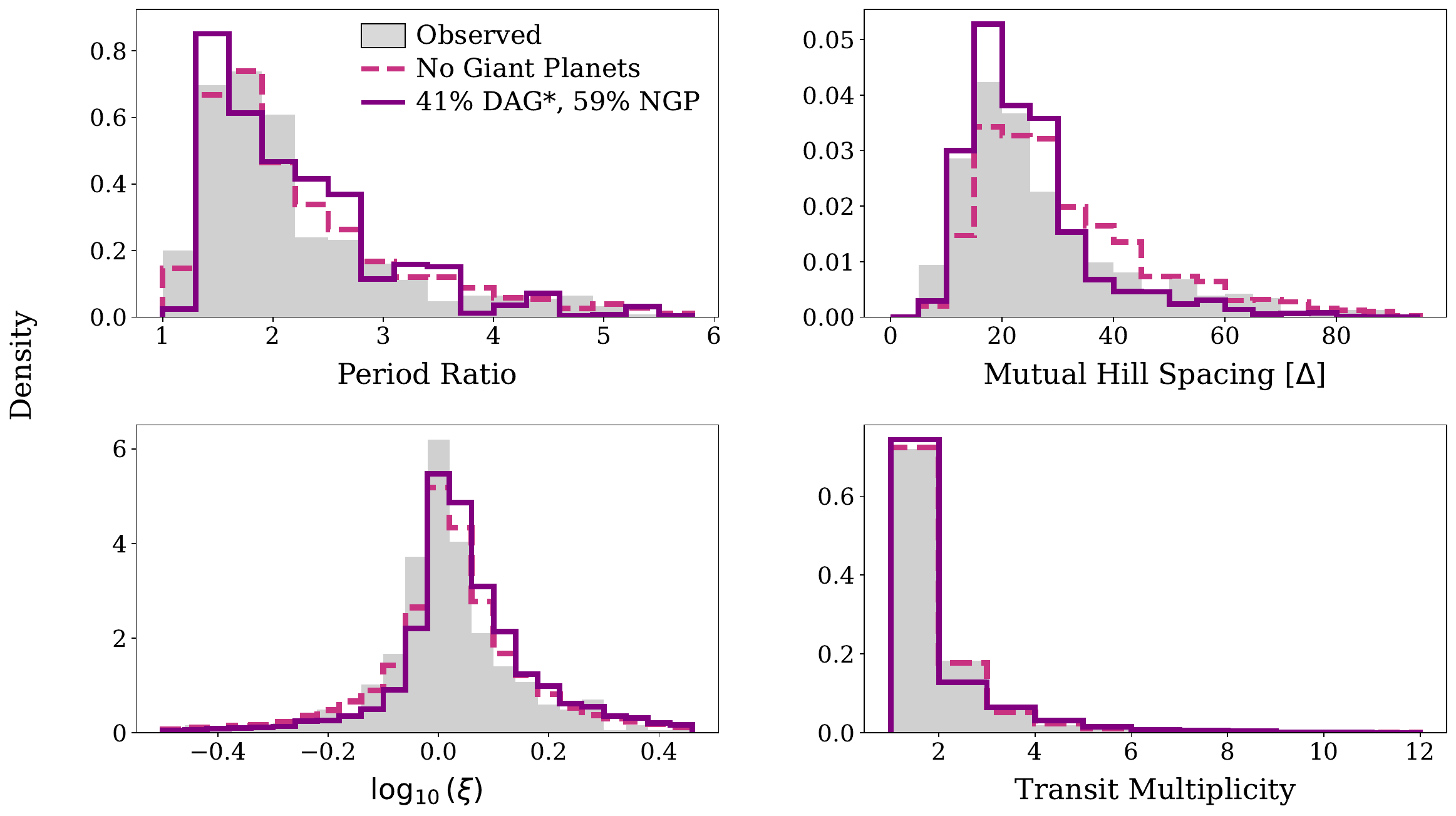} 
\caption{Mock observed period ratio of adjacent planets (top left), Hill spacing between adjacent planets (top right), ratio of transit duration between adjacent planets (bottom left), and the transit multiplicity (bottom right).  The solid gray histogram corresponds to the observed \textit{Kepler} distributions; the dashed pink line shows the simulations without outer giant planets inserted (data from \citealt{MacDonald2020}); and the solid purple shows our mixture with $41\%$ of mock observed systems that are accompanied by outer giants from the DAG* simulations.  With reweighting of underlying systems, the $41\%$ mixture is very similar to the NGP simulations. \label{fig:obs41}}
\end{figure}

\begin{figure}[htbp]
\epsscale{0.85}
\plotone{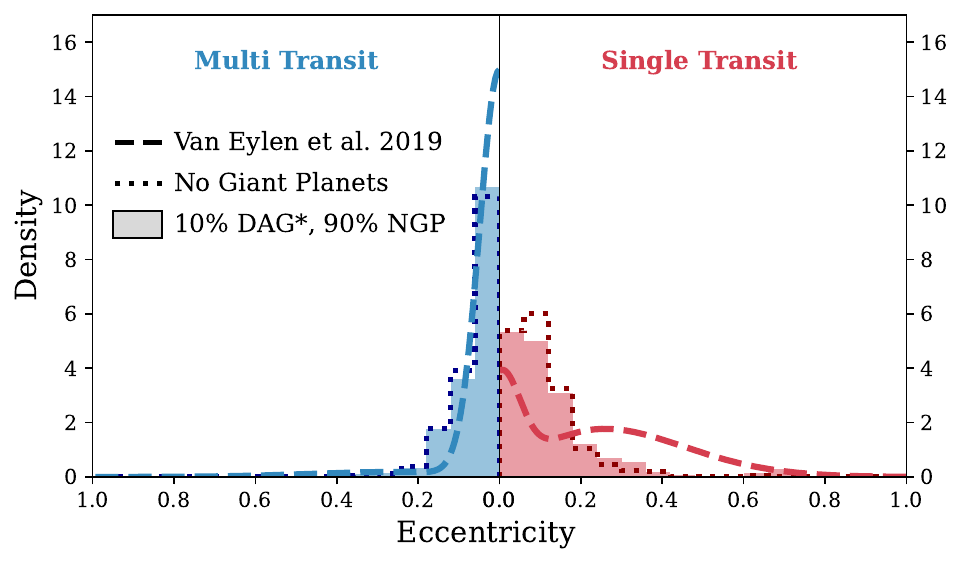} 
\caption{Eccentricity distribution of multi (left, in blue) and singly (right, in red) transiting systems.  The shaded histogram shows our mixture with $10\%$ of mock observed systems that are accompanied by outer giants from the DAG* simulations, the dotted lines shows the NGP simulations, and the inferred observed distribution from \citet{VanEylen2019} plotted as a dashed line. The $10\%$ mixture is not significantly different from the no giant planets case, as may be expected from a low fraction.\label{fig:edist10}}
\end{figure}

\begin{figure}[htbp]
\plotone{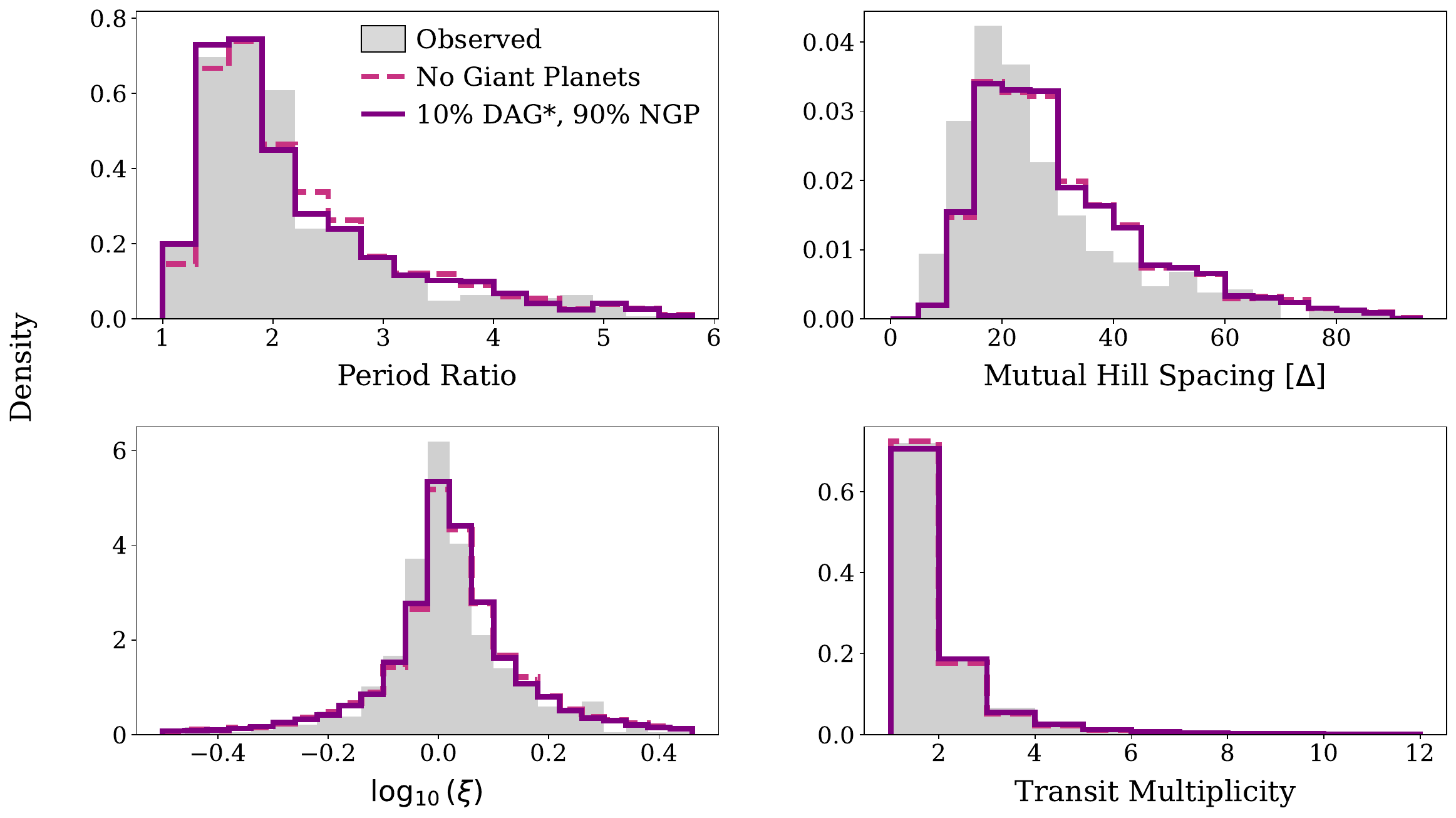} 
\caption{Mock observed period ratio of adjacent planets (top left), Hill spacing between adjacent planets (top right), ratio of transit duration between adjacent planets (bottom left), and the transit multiplicity (bottom right).  The solid gray histogram corresponds to the observed \textit{Kepler} distributions; the dashed pink line shows the simulations without outer giant planets inserted (data from \citealt{MacDonald2020}); and the solid purple line shows our mixture with $10\%$ of mock observed systems that are accompanied by outer giants from the DAG* simulations. With reweighting of underlying systems, the $10\%$ mixture is very similar to the NGP simulations. \label{fig:obs10}}
\end{figure}

In Section \ref{sec:mix}, we explored mixtures of NGP and DAG* simulations.  Figure \ref{fig:obsCompareDAGs} shows a comparison of the observable distributions produced by the DAG and DAG* simulations. Figure \ref{fig:obs41} shows observable properties for the mixture where 41\% of mock observed systems are accompanied by outer giants from the DAG* simulations (Section \ref{sec:mix}). Figure \ref{fig:edist10} shows the eccentricity distribution and Figure \ref{fig:obs10} the observable properties for the mixture where 10\% of mock observed systems are accompanied by outer giants from the DAG* simulations. 

\section{Implementing Gas Damping During the Depleted Gas Disk Stage}
\label{app:gasDamp}

As explained in Section \ref{subsec:simulations}, we integrated our simulations in the presence of a depleted gas disk for 1 Myr.  We incorporated gas damping into our \texttt{REBOUND} \citep{rebound:paper2012, rebound:paper2015a, rebound:paper2015b, rebound:paper2019} simulations using \texttt{REBOUNDx} \citep{reboundx:paper}.  Our prescription for applying the effect of gas damping to all particles in our simulations begins by calculating a damping timescale $\tau$ according to Equation 16 from \citet{DawsonLeeChiang2016}:

\begin{equation}
 \tau= 0.003 d \left(\frac{a}{\text{AU}}\right)^2  \left(\frac{M_\odot}{M_p}\right) \text{yr} \times \left\{
\begin{array}{ll}
       1 & v\leq c_s \\
      \left(\frac{v}{c_s}\right)^3 & v>c_s, i<c_s/v_K \\
      \left(\frac{v}{c_s}\right)^4 & i>c_s/v_K \\
\end{array} 
\right. 
\end{equation}

\noindent where $a$ is the semimajor axis of the particle; $M_p$ is the mass of the particle; $v=\sqrt{e^2 + i^2}v_K$ is the random (epicyclic) velocity; $v_K=na$ is the Keplerian velocity, where $n$ is the body's mean motion; $c_s = 1.29 \text{ km/s} \left(a/\text{AU} \right)^{-1/4}$ is the gas sound speed; and $d$ is the gas depletion factor.  $d=1$ corresponds to a gas surface density $\Sigma_g = 1700 \text{ g cm}^{-2}$ at 1 AU, which approximates the gas density of the full minimum mass solar nebula. $d>1$ denotes a more depleted nebula.

This timescale $\tau$ is used to damp both the particles' eccentricities and inclinations, described by \citep{KominamiIda2002_Icar}:

\begin{equation}
    \begin{split}
        \dot{e}/e &= -1/\tau \\
        \dot{i}/i &= -1/(2\tau)
    \end{split}
\end{equation}

\noindent which implies $\tau = \tau_e = \tau_i/2$, where $\tau_e$ and $\tau_i$ are the particle's eccentricity and inclination timescales, respectively.  Following from previous iterations of implementing gas damping forces, our simulations were run with $\tau_i$ incorrectly set to $\tau/2$.  We performed a suite of simulations to check the population-level effects of this difference in timescale coefficients.  We found no significant differences in any properties, except the corrected inclinations were larger. However, the inclinations in both cases were so small that they were still consistent with zero.  Additionally, we anticipate that these minuscule differences in inclination would be erased within our simulations after undergoing the subsequent 27 Myr of post-gas evolution.   

We use these damping timescales to calculate the acceleration $\vec{a}_\text{damp}$ \citep{PapaloizouLarwood2000_DampTimescale}:

\begin{equation}
    \vec{a}_\text{damp} = -\frac{2 (\vec{v} \cdot \vec{r}) \hat{r}}{r^2 \tau_e} - \frac{2 (\vec{v} \cdot \hat{z}) \hat{z}}{\tau_i}
\end{equation}

\noindent where the velocity vector $\vec{v} = (v_x, v_y, v_z)$; the position vector $\vec{r} = (x, y, z)$; and $\hat{z}$ is the unit vector in the $z$ (vertical) direction.

This prescription for incorporating gas damping timescales is available as \texttt{gas\_damping\_timescale} within the package \texttt{REBOUNDx} at: \url{https://github.com/dtamayo/reboundx}.

\clearpage
\bibliography{refs}{}
\bibliographystyle{aasjournal}

\end{document}